\documentclass[
  aps,
  physrev,
  twocolumn,
  11pt,
  tightenlines,
  superscriptaddress,
  amsmath,
  amssymb,
  floatfix
]{revtex4-2}

\usepackage{mathtools}
\usepackage{amsthm}
\usepackage{graphicx}
\usepackage{booktabs}
\usepackage{bm}
\usepackage{braket}
\usepackage[hidelinks]{hyperref}
\hypersetup{
  pdftitle={Quantum Sensing of Non-Repeatable Events Enhanced by In-Sensor Quantum Reservoir Computing},
  pdfauthor={Daiki Sasaki; Hideaki Kawaguchi; Yuichiro Matsuzaki}
}
\usepackage{placeins}

\theoremstyle{plain}
\newtheorem{proposition}{Proposition}[section]

\newcommand{\task}{\mathrm{task}}
\newcommand{\Acc}{\mathrm{Acc}}
\newcommand{\Amin}{A_{\min}}
\newcommand{\DD}{\mathrm{DD}}
\newcommand{\Reservoir}{\mathrm{R}}
\newcommand{\Ncand}{N_{\mathrm{cand}}}
\newcommand{\mR}{m_{\mathrm{R}}}
\newcommand{\Law}{\operatorname{Law}}

\begin{document}

\title{Quantum Sensing of Non-Repeatable Events Enhanced by In-Sensor \newline Quantum Reservoir Computing}

\author{Daiki Sasaki}
\affiliation{Department of Electrical, Electronic, and Communication Engineering, Faculty of Science and Engineering, Chuo University, Tokyo, Japan}
\affiliation{Department of Engineering Science, The University of Electro-Communications, Chofu, Tokyo 182-8585, Japan}
\author{Hideaki Kawaguchi}
\affiliation{Medical Research Center for Pre-Disease State (Mebyo) AI, Graduate School of Medicine, The University of Tokyo, 7-3-1, Hongo, Bunkyo-ku, Tokyo 113-0033, Japan}
\author{Yuichiro Matsuzaki}
\affiliation{Department of Electrical, Electronic, and Communication Engineering, Faculty of Science and Engineering, Chuo University, Tokyo, Japan}

\begin{abstract}
High-density nitrogen-vacancy (NV) ensembles in diamond enable sensitive magnetometry. Many quantum-sensing protocols rely on reproducible target fields, allowing repeated measurements under different sensing conditions. In dynamical decoupling (DD) magnetometry, sweeping the interval between $\pi$ pulses across repeated measurements enables estimation of the frequency and amplitude of an unknown alternating field. For non-repeatable events, however, the same field waveform cannot be reproduced for measurements under different settings. Here we propose NV-based in-sensor quantum reservoir computing (NV-QRC) for sensing such events. Field-driven many-body dynamics and simultaneous fluorescence readout from multiple spatial regions provide a classical classifier with multiple features from a single event. For binary phase classification, we benchmark NV-QRC against DD magnetometry using a single pulse sequence fixed in advance. We show that NV-QRC can retain class-dependent information in regimes where the fixed-DD protocol fails to capture it. These results identify non-repeatable-event sensing as a promising application of quantum reservoir computing.
\end{abstract}

\maketitle

\section{Introduction}
\label{sec:introduction}

Quantum sensing detects weak magnetic and electric fields through the evolution and measurement of a quantum system. Nitrogen-vacancy (NV) centers in diamond, representative solid-state quantum sensors, can be optically initialized and read out through fluorescence and controlled with microwaves at room temperature, enabling local-field sensing with high spatial resolution~\cite{taylor2008highSensitivity,rondin2014magnetometry,degen2017quantum,barry2020sensitivity,santagati2019magnetic}. In NV ensembles, spatially resolved fluorescence can be acquired in parallel, and increasing the NV density increases the number of NV centers and fluorescence photons within a small readout region~\cite{pham2011magneticImaging,barGill2012suppression,acosta2009highDensity,barry2020sensitivity}. At the same time, in high-density NV ensembles, inhomogeneity in transverse strain and local electric fields produces inhomogeneous spectral broadening, while dipolar interactions between NV centers can limit the coherence time; both effects can degrade sensing performance~\cite{stanwix2010coherence, matsuzakiImproving2015,matsuzakiOpticallyDetectedMagnetic2016,bauch2020decoherence}. The control sequence used for NV magnetometry also depends on the temporal variation of the target field. In dynamical decoupling (DD), a representative control method for alternating-current magnetometry, the interval between $\pi$ pulses determines the frequency-selective phase response~\cite{deLange2010universalDD,deLange2011singleSpin,pham2012enhancedMultispin,rondin2014magnetometry,barry2020sensitivity}.

Many quantum-sensing methods implicitly assume that the target field is reproducible and can be measured repeatedly under the same conditions. For example, in alternating-current magnetometry using DD, if the same field signal can be measured repeatedly, the frequency and amplitude of a field with unknown frequency can be estimated by sweeping the interval between the $\pi$ pulses. In contrast, when the target is a non-repeatable event, repeated measurements under different conditions are unavailable. Here, we define a non-repeatable event as one for which the same time-dependent signal cannot be reproduced and remeasured under a different measurement condition. Accurately extracting information from short-duration, weak time-dependent signals arising from such non-repeatable events is therefore an important challenge in quantum sensing when measurement opportunities are limited.

Quantum reservoir computing (QRC) has been studied as a framework for using the response of a quantum system driven by a time-varying input to generate features for machine learning. Reservoir computing (RC) uses fixed input-driven dynamics to generate features and typically trains only the output layer~\cite{jaeger2001echo,maass2002realTime,lukosevicius2009reservoir,tanaka2019recent}. In QRC, the time evolution of a quantum system implements the fixed dynamics, while measurements of the resulting state provide features for the classical output layer~\cite{fujii2017harnessing,nakajima2019boosting,yasudaRepeatedMeasurementsQRC2023,mujalTimeSeriesMeasurements2023,zhuContinuousMeasurementQRC2025,tranHigherOrderQRC2020}. This makes it possible to use the high-dimensional state space and many-body dynamics of a quantum system as computational resources that transform a time-dependent input into diverse measurement responses. Despite this possibility, the problem settings in which QRC is useful remain unclear. In particular, when QRC processes pre-acquired classical data, its advantages over other information-processing models under the same input information and available resources have not been firmly established~\cite{mujal2021opportunities,cerezoChallengesOpportunitiesQML2022}.

Recent studies have directly applied continuous-time physical signals to quantum systems and constructed features from the resulting measurements, forming a connection between quantum sensing and QRC\@. Senanian \textit{et al.} demonstrated an analog QRC in which microwave signals are applied to a superconducting quantum circuit and classified using measurements taken at multiple times~\cite{senanianMicrowaveSignalProcessing2024}. Settino \textit{et al.} numerically evaluated the magnetic response of a superconducting-flux-qubit network and its time-series processing performance as a QRC~\cite{settino2026topologyEnhanced}. However, it remains unclear whether using features derived from a quantum response offers an advantage over standard quantum sensing methods for the same signal-classification task. Whether integrating a quantum sensor with QRC produces a classification benefit unavailable to standard quantum sensing therefore remains an important open question.

We propose NV-QRC, which uses a high-density NV center ensemble as both a quantum sensor and a physical reservoir, and consider the classification of magnetic-field waveforms associated with non-repeatable events. In the sensing role, we assume wide-field magnetometry with a high-density NV center ensemble; in the reservoir role, we use interactions between NV centers. We compare this method with standard alternating-current magnetometry using DD\@. The method does not feed pre-acquired classical data into a quantum system; instead, the target field directly drives the NV ensemble, whose response provides the classification features. In NV-QRC, no additional microwave control is applied during the interaction with the field; instead, the spatially resolved spin responses produced by the input field and the internal Hamiltonian, including transverse strain and NV--NV dipolar interactions, are read out in parallel as fluorescence after the field interaction. This provides multiple features without measuring the same event again.

For binary phase classification of sine and cosine waves with event-dependent amplitudes, we compare DD using a single Carr--Purcell--Meiboom--Gill (CPMG) pulse sequence common to all events with NV-QRC including dipolar interactions between NV centers. For both the DD features and the NV-QRC features obtained by setting only the interaction term to zero, there are conditions under which the class-conditional feature distributions coincide and the classes cannot be discriminated from those features alone. Under the numerical conditions examined, NV-QRC with dipolar interactions retains class-dependent differences in a feature vector composed of responses from multiple regions even under those conditions, enabling classification of non-repeatable events.

These results extend the scope of QRC from processing pre-acquired classical data with a quantum system to using the acquisition process of a physical signal itself for feature generation. By assigning the roles of sensor and physical reservoir to the same quantum system, this method provides a sensing-oriented application of QRC for non-repeatable events.

\section{Background and preliminaries}
\label{sec:background}

\subsection{Sequence classification with reservoir computing}
\label{sec:background_rc}

Reservoir computing (RC) is a machine-learning framework that separates fixed dynamics driven by a time-series input from a trainable output layer that maps the resulting state to a target variable. The framework originated with the echo state network (ESN) and the liquid state machine (LSM), whose common structure was later unified under the term reservoir computing~\cite{jaeger2001echo,jaeger2004harnessing,maass2002realTime,verstraeten2007unification,lukosevicius2009reservoir}. In basic RC, the input coupling and internal dynamics are fixed, and only the output layer is trained for a given task.

Let $\nu$ index the samples. At discrete time $k=1,\ldots,K$, let the input for sample $\nu$ be $\bm u_{\nu,k}\in\mathbb R^{d_{\mathrm{in}}}$ and the reservoir state be $\bm h_{\nu,k}\in\mathbb R^d$. An ESN, a standard discrete-time RC model, can, for example, be written for $k=0,\ldots,K-1$ as
\begin{equation}
\begin{aligned}
\bm h_{\nu,k+1}
&=
\tanh\!\left(
\bm W_{\mathrm{in}}\bm u_{\nu,k+1}
+\bm W\bm h_{\nu,k}
\right),
\\
\bm h_{\nu,0}&=\bm h_0
\end{aligned}
\label{eq:generic_discrete_rc}
\end{equation}
Here $\tanh$ acts elementwise, $\bm W_{\mathrm{in}}\in\mathbb R^{d\times d_{\mathrm{in}}}$ is the input coupling, $\bm W\in\mathbb R^{d\times d}$ is the internal coupling, and $\bm h_0\in\mathbb R^d$ is an initial state shared across samples; all are fixed during training. The state $\bm h_{\nu,k}$ reflects the input history up to time $k$. In sequence classification, where one label is assigned to an input sequence of length $K$, the terminal state $\bm h_{\nu,K}$ is passed to a downstream classifier as a feature representing $\{\bm{u}_{\nu,k}\}_{k=1}^K$. For example, in a linear classifier for binary classification, the mean and standard deviation of each component of the terminal reservoir states obtained from the training data are first computed and used for standardization. Let the mean and standard deviation of component $a$ be $\overline h_a\in\mathbb R$ and $\sigma_a\in\mathbb R_{>0}$, respectively, and define the $a$-th component of the standardized state $\widetilde{\bm h}_{\nu,K}\in\mathbb R^d$ by
\begin{equation}
\widetilde h_{\nu,K,a}
:=
\frac{h_{\nu,K,a}-\overline h_a}{\sigma_a},
\qquad
a=1,\ldots,d
\label{eq:generic_rc_standardization}
\end{equation}
Let the true binary label of sample $\nu$ be $Y_\nu\in\{0,1\}$, and define the signed target used by the ridge classifier as
\begin{equation}
t_\nu
:=
2Y_\nu-1
=
\begin{cases}
-1, & Y_\nu=0,\\
+1, & Y_\nu=1
\end{cases}
\label{eq:generic_rc_signed_target}
\end{equation}
In the ridge classifier, the intercept $b\in\mathbb R$ and coefficient vector $\bm\beta\in\mathbb R^d$ linearly predict the signed target $t_\nu$ from the standardized reservoir state and are obtained by minimizing
\begin{equation}
\begin{aligned}
\left(\widehat b,\widehat{\bm\beta}\right)
&:=
\operatorname*{arg\,min}_{b,\bm\beta}
\biggl\{
\sum_{\nu\in\mathcal I_{\mathrm{tr}}}
\left(
t_\nu-b-\bm\beta^{\mathsf T}\widetilde{\bm h}_{\nu,K}
\right)^2\\
&\hspace{8em}
{}+
\alpha_{\mathrm{reg}}\lVert\bm\beta\rVert_2^2
\biggr\}
\end{aligned}
\label{eq:generic_rc_ridge_training}
\end{equation}
Here $\mathcal I_{\mathrm{tr}}$ is the index set of training samples and $\alpha_{\mathrm{reg}}\ge0$ is the regularization coefficient. The decision score and predicted label after training are
\begin{equation}
\widehat y_\nu
=
\widehat b
+
\widehat{\bm\beta}^{\mathsf T}\widetilde{\bm h}_{\nu,K},
\qquad
\widehat Y_\nu
:=
\begin{cases}
1, & \widehat y_\nu>0,\\
0, & \widehat y_\nu\le0
\end{cases}
\label{eq:generic_rc_classifier}
\end{equation}
Thus, $\widehat y_\nu$ is a real-valued prediction of the signed target $t_\nu$, whose sign is converted to the binary label $\widehat Y_\nu$. The output-layer weight
\begin{equation}
\bm w_{\mathrm{out}}
:=
\left(
\widehat b,
\widehat{\bm\beta}^{\mathsf T}
\right)^{\mathsf T}
\in\mathbb R^{d+1}
\label{eq:generic_rc_output_weight}
\end{equation}
is learned from labeled training data. The implementation used in this work is described in Sec.~\ref{sec:numerical_setup}.

For a continuous-time physical reservoir, let the input at time $t$ be $\bm u_\nu(t)\in\mathbb R^{d_{\mathrm{in}}}$ and the reservoir state be $\bm h_\nu(t)\in\mathbb R^d$. The corresponding state evolution can be written as
\begin{equation}
\begin{aligned}
\frac{d\bm h_\nu(t)}{dt}
&=
\bm G\!\left(\bm h_\nu(t),\bm u_\nu(t)\right),
\\
\bm h_\nu(0)&=\bm h_0,
\qquad 0\le t\le T
\end{aligned}
\label{eq:generic_continuous_rc}
\end{equation}
where $\bm G:\mathbb R^d\times\mathbb R^{d_{\mathrm{in}}}\to\mathbb R^d$ defines the fixed input-driven dynamics and $\bm h_0\in\mathbb R^d$ is the initial state. In physical RC, this state evolution is provided not by an artificial neural network but by physical dynamics such as optical, electronic, mechanical, or spin systems~\cite{tanaka2019recent,nakajima2020physical}. A principal advantage of RC and physical RC is that training can be restricted to the output layer rather than using backpropagation to train the entire internal dynamics.

As an example of sequence classification with one label per input sequence, Bianchi \textit{et al.} evaluated the use of the terminal reservoir state of a multivariate time series as a vector representation supplied to a classifier~\cite{bianchi2021reservoir}. In physical RC, Du \textit{et al.} applied finite pulse sequences to a dynamical memristor and used the measured device state after the pulse sequence for classification~\cite{du2017memristor}. Reservoirs are also commonly evaluated using nonlinear autoregressive moving average (NARMA) prediction and short-term memory (STM) benchmarks, which measure prediction performance at each time or the ability to reproduce input history. The echo state property (ESP), in which the influence of different initial states under the same input history vanishes with time, is another dynamical property discussed for reservoirs~\cite{lukosevicius2009reservoir}. Because this work considers sequence classification with one label per input sequence, we do not evaluate NARMA prediction, the STM benchmark, or the ESP.

\subsection{Quantum reservoir computing}
\label{sec:background_qrc}

Quantum reservoir computing (QRC) uses the input-driven evolution of a quantum system as a reservoir and obtains features by measuring the resulting state for a downstream classical output layer. Fujii and Nakajima proposed QRC based on the dynamics of a disordered quantum many-body system for time-series information processing; subsequent work has explored spatial multiplexing and a variety of quantum systems, inputs, and tasks~\cite{fujii2017harnessing,nakajima2019boosting, yasudaRepeatedMeasurementsQRC2023,mujalTimeSeriesMeasurements2023,zhuContinuousMeasurementQRC2025, tranHigherOrderQRC2020,mujal2021opportunities}.

For sample $\nu$, let the input at time $t$ be $\bm u_\nu(t)\in\mathbb R^{d_{\mathrm{in}}}$, the quantum state on a Hilbert space $\mathcal H$ be $\rho_\nu(t)$, and the observation vector at the end of the input be $\bm h_\nu(T)\in\mathbb R^d$. The quantum evolution and observables are then
\begin{align}
\frac{d\rho_\nu(t)}{dt}
&=
\mathcal L_{\bm u_\nu(t)}[\rho_\nu(t)],
\qquad
\rho_\nu(0)=\rho_0,
\label{eq:generic_qrc_dynamics}
\\
\bm h_\nu(T)
&:=
\bigl(
\operatorname{Tr}[\rho_\nu(T)O_1],\ldots,
\operatorname{Tr}[\rho_\nu(T)O_d]
\bigr)^{\mathsf T}
\label{eq:generic_qrc_features}
\end{align}
where $\mathcal L_{\bm u_\nu(t)}$ is a fixed generator of time evolution that depends on the input, $\rho_0$ is an initial state on $\mathcal H$ shared across samples, and $O_1,\ldots,O_d$ are fixed observables on $\mathcal H$. For a closed quantum system with a scalar input $u(t)\in\mathbb R$, for example, $\mathcal L_{u(t)}[\rho]=-\mathrm i[H_0+u(t)H_{\mathrm{in}},\rho]/\hbar$, where $H_0$ is the input-independent internal Hamiltonian and $H_{\mathrm{in}}$ is the Hamiltonian coupled to the input. The observation vector $\bm h_\nu(T)$ is passed to the classifier introduced in the preceding subsection as a feature representing the input sequence. Thus, quantum dynamics and quantum measurement implement the state transformation and feature acquisition of RC\@. QRC also includes constructions using repeated input injection, measurements at multiple times, and dissipation.

Analog QRC has also been demonstrated in settings where a continuous-time physical signal acts directly on a quantum system and measurements of its response provide classification features~\cite{senanianMicrowaveSignalProcessing2024}. We consider this setting and use observations at the end of the input as classification features. The specific quantum system and readout method are described below.

\subsection{Feature distributions and limits of binary classification}
\label{sec:background_classification}

Regardless of the physical system, measurement method, or reservoir type, the information available to a classifier is constrained by the probability distributions of the acquired features. Let $P_0$ and $P_1$ be the feature distributions for two classes. Define the total variation distance between the two distributions on feature space by
\begin{equation}
\lVert P_0-P_1\rVert_{\mathrm{TV}}
:=
\sup_{\mathcal A}|P_0(\mathcal A)-P_1(\mathcal A)|
\label{eq:tv_definition}
\end{equation}
where $\mathcal A$ is any event in feature space. When the two class priors are equal, the Bayes-optimal accuracy based only on these features is
\begin{equation}
\Acc^\ast(P_0,P_1)
:=
\frac12\left(1+\lVert P_0-P_1\rVert_{\mathrm{TV}}\right)
\label{eq:bayes_limit}
\end{equation}
\cite{tsybakov2009introduction}. Thus, if $P_0=P_1$, the Bayes-optimal accuracy based on the acquired features is $1/2$. In later sections, Eq.~\eqref{eq:bayes_limit} is applied to the class-conditional feature distributions produced by each measurement protocol.

\subsection{NV centers and NV ensembles}
\label{sec:background_nv}
The negatively charged nitrogen-vacancy (NV) center in diamond is a point defect consisting of a substitutional nitrogen atom adjacent to a vacancy, and its electronic ground state is a spin-1 triplet. Taking the NV axis to be the $z$ axis, the Hamiltonian for the electronic spin degrees of freedom is
\begin{equation}
\begin{split}
H_{\mathrm{NV}}
={}&D_{\mathrm{gs}}S_z^2
+\gamma_e\bm B_{\mathrm{ext}}\!\cdot\!\bm S\\
&+E_x(S_x^2-S_y^2)
+E_y(S_xS_y+S_yS_x)
\end{split}
\label{eq:standard_nv_hamiltonian}
\end{equation}
\cite{rondin2014magnetometry,degen2017quantum,barry2020sensitivity}. Here $\bm S=(S_x,S_y,S_z)$ is the spin-1 operator, $D_{\mathrm{gs}}/2\pi\simeq2.87\,\mathrm{GHz}$ is the zero-field splitting between $m_s=0$ and $m_s=\pm1$, $\gamma_e$ is the magnitude of the electron-spin gyromagnetic ratio, and $\bm B_{\mathrm{ext}}$ is the external magnetic field. The effective coefficients $E_x$ and $E_y$ describe the coupling between $m_s=\pm1$ induced by transverse crystal strain or a local electric field.

NV magnetometry combines electronic-spin initialization, microwave state manipulation, and fluorescence readout of the spin state. Under green-light illumination, the initial fluorescence intensity depends on whether the electronic spin was in $m_s=0$ or $m_s=\pm1$ before illumination, so the spin state can be read out from the fluorescence. Continuing the illumination optically pumps the electronic spin predominantly into $m_s=0$. A typical measurement cycle begins with green-light initialization, followed by interaction with the target field with any required microwave operations, and ends with green-light fluorescence detection. A microwave magnetic field with a component transverse to the NV axis drives transitions between $m_s=0$ and $m_s=\pm1$ when its frequency is resonant with the spin transition. Microwaves can be applied as a continuous wave or as pulses for detecting electron-spin resonance, preparing the spin state, controlling the state during interaction with the target field, and transforming the state before readout~\cite{rondin2014magnetometry,degen2017quantum,barry2020sensitivity}.

These initialization, state-manipulation, and fluorescence-readout procedures apply to both individual NV centers and NV ensembles. An individual near-surface NV center can measure a local field with nanometer-scale spatial resolution by positioning the sensor close to the sample~\cite{taylor2008highSensitivity}. However, the information obtained from a single NV center is limited. Measurements with NV ensembles improve sensitivity by acquiring fluorescence from many NV centers simultaneously, while spatially resolving the fluorescence image with a camera enables parallel measurements of fields associated with multiple detection regions~\cite{pham2011magneticImaging}. In typical wide-field magnetometry with an NV ensemble, green light illuminates the full field of view, the same microwave pulses are applied to the NV centers across that field, and a camera acquires a spatially resolved fluorescence image~\cite{pham2011magneticImaging,barGill2012suppression}.

The main experimental motivation for using a high-density NV ensemble is to ensure sufficient numbers of NV centers and fluorescence photons even in small, spatially resolved detection regions, thereby improving sensitivity while retaining spatial resolution. Sensitivity, however, depends not only on the number of NV centers but also on the linewidth, coherence time, and optical contrast~\cite{acosta2009highDensity,barry2020sensitivity}. In addition, sample-processing conditions and material properties make the transverse strain and local electric field experienced by each NV center different, producing distributions of $E_x$ and $E_y$ in Eq.~\eqref{eq:standard_nv_hamiltonian}~\cite{stanwix2010coherence, Zhu2014DarkStates, matsuzakiImproving2015,matsuzakiOpticallyDetectedMagnetic2016,bauch2020decoherence}.
As the NV concentration increases, the mean distance between NV centers decreases, and the magnetic dipolar interaction, which scales inversely with the cube of the distance, affects the quantum dynamics. This is physically distinct from the inhomogeneity in transverse strain arising from sample fabrication. NV ensembles with strong dipolar interactions have been realized, and their many-body dynamics and applications to quantum metrology have been studied~\cite{choi2020robustHamiltonian,hughes2025stronglyInteracting,zhouQuantumMetrologyStrongly2020}. Gao \textit{et al.} showed that the intrinsic dipolar interactions of an NV ensemble can be modified by encoding qubits in dressed states and used for alternating-current magnetometry~\cite{gao2026dressedState}. Thus, although inhomogeneity and interactions in high-density NV ensembles can limit coherence, depending on the state-preparation and readout methods they can also form part of the quantum dynamics that transform an input field into distinct measurement outcomes.

The two main fluorescence-based methods for detecting NV electron-spin resonance are continuous-wave optically detected magnetic resonance (cw-ODMR) and pulsed ODMR\@. In cw-ODMR, the microwave frequency is swept while green light and microwaves are applied continuously, and the reduction in fluorescence intensity at the resonance frequency is detected. This method has a relatively simple measurement system and control procedure and facilitates the acquisition of resonance spectra over a broad frequency range. By contrast, because optical spin reinitialization and microwave driving of the transition occur simultaneously, the resonance contrast is limited. The laser and microwave intensities can also broaden the linewidth and thereby reduce the frequency resolution and magnetic-field sensitivity~\cite{rondin2014magnetometry,barry2020sensitivity}.

In pulsed ODMR, by contrast, green-light initialization, microwave-pulse state manipulation, and green-light fluorescence readout are separated in time. This avoids optical pumping during microwave manipulation, suppresses linewidth broadening relative to cw-ODMR, and facilitates high resonance contrast and frequency resolution. Designing the timing of the microwave pulses also enables coherent state manipulation~\cite{degen2017quantum,pham2012enhancedMultispin}. The pulsed-ODMR procedure is more complex than the cw-ODMR procedure because the timing and intensity of the laser and microwave pulses must be controlled, and errors in these quantities affect the measurement~\cite{rondin2014magnetometry,barry2020sensitivity}.
The next subsection describes dynamical decoupling as a representative method for shaping the response to an alternating-current magnetic field through microwave control.
\subsection{Alternating-current magnetometry with dynamical decoupling}
\label{sec:background_dd}

Dynamical decoupling (DD) applies control pulses to a quantum system to reverse the sign of phase accumulation caused by environmental noise and average out noise that varies more slowly than the control. Using the same modulation for magnetometry enables selective phase accumulation from alternating-current field components that are synchronized with the pulse interval. DD is therefore used both to preserve coherence and to perform frequency-selective measurements of alternating-current fields~\cite{deLange2010universalDD,deLange2011singleSpin,degen2017quantum}.

In a representative configuration for alternating-current magnetometry with a DD sequence applied to an NV center, a static bias field is applied along the NV axis to lift the degeneracy of $m_s=\pm1$. The transition frequencies of $m_s=0\leftrightarrow m_s=+1$ and $m_s=0\leftrightarrow m_s=-1$ then separate, so one transition can be selectively driven with resonant microwaves and the selected two levels can be treated as an effective spin-1/2 system~\cite{rondin2014magnetometry,degen2017quantum,barry2020sensitivity}. The DD-based alternating-current magnetometry used as the comparison method in this work also relies on this effective two-level description.

The procedure for measuring an alternating-current field with DD using an NV center consists of optical initialization, microwave state preparation, interaction with the target field, microwave manipulation before readout, and fluorescence readout. First, green light initializes the electronic spin predominantly in $m_s=0$, and a resonant microwave $\pi/2$ pulse prepares a superposition state. The spin is then exposed to the target field during the sensing interval $0\le t\le T$, while multiple resonant microwave $\pi$ pulses are applied. Finally, a $\pi/2$ pulse before readout converts the accumulated phase into a level-occupation probability, and green light is applied to detect spin-dependent fluorescence~\cite{rondin2014magnetometry,barry2020sensitivity}. The relation between the state preparation, evolution under the CPMG sequence, and readout used in this work is shown mathematically in Appendix~\ref{app:dd_readout}.

The $\pi/2$ and $\pi$ pulses are implemented by appropriately setting the intensity and duration of microwaves resonant with the spin transition. Below, we consider ideal instantaneous pulses whose widths can be neglected. In the Carr--Purcell--Meiboom--Gill (CPMG) sequence used here, $N_p$ ideal $\pi$ pulses are applied during the sensing time $T$ at
\begin{equation}
t_k
:=
\left(k-\frac12\right)\frac{T}{N_p},
\qquad
k=1,\ldots,N_p
\label{eq:cpmg_pulse_times}
\end{equation}
The corresponding DD frequency is
\begin{equation}
f_{\DD}
:=
\frac{N_p}{2T}
\label{eq:fdd_cpmg}
\end{equation}
Let $t_0:=0$ and $t_{N_p+1}:=T$. The phase modulation generated by the CPMG sequence is
\begin{equation}
\begin{aligned}
&\eta_{N_p}(t)
:=
(-1)^k,\\
&
t_k\le t<t_{k+1},
\qquad
k=0,\ldots,N_p
\end{aligned}
\label{eq:cpmg_toggling}
\end{equation}
Here $N_p$ and $T$ specify the CPMG sequence, and $f_{\DD}$ is determined by them.

The phase accumulated by a field $B(t)$ along the NV axis up to time $T$ is
\begin{equation}
\varphi(T)
=
\gamma_e\int_0^T\eta_{N_p}(t)B(t)\,dt
\label{eq:generic_dd_phase}
\end{equation}
The Fourier transform of the toggling function is
\begin{equation}
\widetilde\eta_{N_p}(\Omega)
:=
\int_0^T\eta_{N_p}(t)e^{\mathrm i\Omega t}\,dt
\label{eq:generic_dd_filter}
\end{equation}
where $\Omega\in\mathbb R$ is an angular frequency. The complex quantity $\widetilde\eta_{N_p}(\Omega)$ weights the contribution of the magnetic-field component at angular frequency $\Omega$ to the accumulated phase, and its magnitude represents the sensitivity to that component. For CPMG, the main sensitivity band appears near $f_{\DD}$. Thus, the phase obtained from one fixed CPMG sequence is a single weighted integral of the field waveform, and obtaining additional information with a different pulse interval generally requires another measurement cycle.

In one projective measurement of an individual NV center, one of the eigenvalues of the spin observable is obtained stochastically, so its expectation value cannot be determined from a single measurement. Estimating a spin expectation value with an individual NV center therefore requires repeating the same state preparation, control, and readout many times and averaging the outcomes. In an actual fluorescence readout, the number of photons obtained from an individual NV in one readout is also limited, producing statistical fluctuations associated with photon detection. For $M$ independent repetitions, these statistical errors typically decrease in proportion to $M^{-1/2}$~\cite{degen2017quantum,barry2020sensitivity}.

In an NV ensemble, many NV centers within one detection region can be prepared and controlled under the same conditions, and their fluorescence can be acquired simultaneously. Each NV center in the ensemble then effectively provides one independent realization of the measurement. Therefore, if a detection region contains sufficiently many NV centers and their collective fluorescence can be detected with a high signal-to-noise ratio, the spin expectation value can in principle be estimated from the ensemble average obtained in a single measurement cycle~\cite{pham2011magneticImaging,barry2020sensitivity}.
Based on these considerations, the next section defines an event as one realization of the magnetic-field waveform during the sensing interval. It then formulates the classification of non-repeatable events using a single measurement protocol fixed in advance for all events; the same waveform cannot be remeasured under different conditions.

\section{Problem setting}
\label{sec:problem}
\subsection{Classification of non-repeatable events with a fixed measurement protocol}

Let $T$ denote the sensing time. We call one realization of the magnetic-field waveform acting on the quantum sensor during $0\le t\le T$ an event and distinguish events by the sample index $\nu$.
In this work, classification of non-repeatable events refers to a setting in which the class of each event must be determined even though the same physical magnetic-field waveform cannot be reproduced and measured again under a different measurement condition.

In a quantum sensor, information is encoded in the quantum state through interaction with the magnetic field and is then read out by measuring a prescribed observable. In general, a measurement projects or reinitializes the quantum state, so the pre-measurement state is lost and the same state cannot be measured again with another control sequence or observable. If the target waveform itself is not reproducible, it is also impossible to prepare a new quantum state and measure the same event again. Thus, the information acquired from each event is limited to that obtained with the state preparation, control sequence, and readout operation fixed in advance and used for all events.

Under this non-repeatability constraint, one measurement protocol independent of the class label, the individual waveform, and the task frequency introduced below is fixed in advance for all events. The measurement protocol includes state preparation, control during sensing, operations before readout, the observable, and the set of NV-QRC readout regions defined later in Sec.~\ref{sec:nv_qrc_protocol}.

All events in both the training and test sets are measured using this same fixed protocol. A classical classifier is then trained on the measurements and labels of the training events and used to infer the labels of the test events, whose labels are not provided to the classifier. We assume that the training and test events follow the same class-conditional distributions and experimental conditions. In this setting, a downstream classifier cannot recover information that was not acquired by the fixed measurement. The ability to classify the events is therefore constrained not only by the expressive power of the classical classifier but also by what information about the waveform the fixed measurement protocol retains in the measured features.
\subsection{Assumed measurement platform based on an NV ensemble}

\begin{figure*}[t]
\centering
\includegraphics[width=\textwidth]{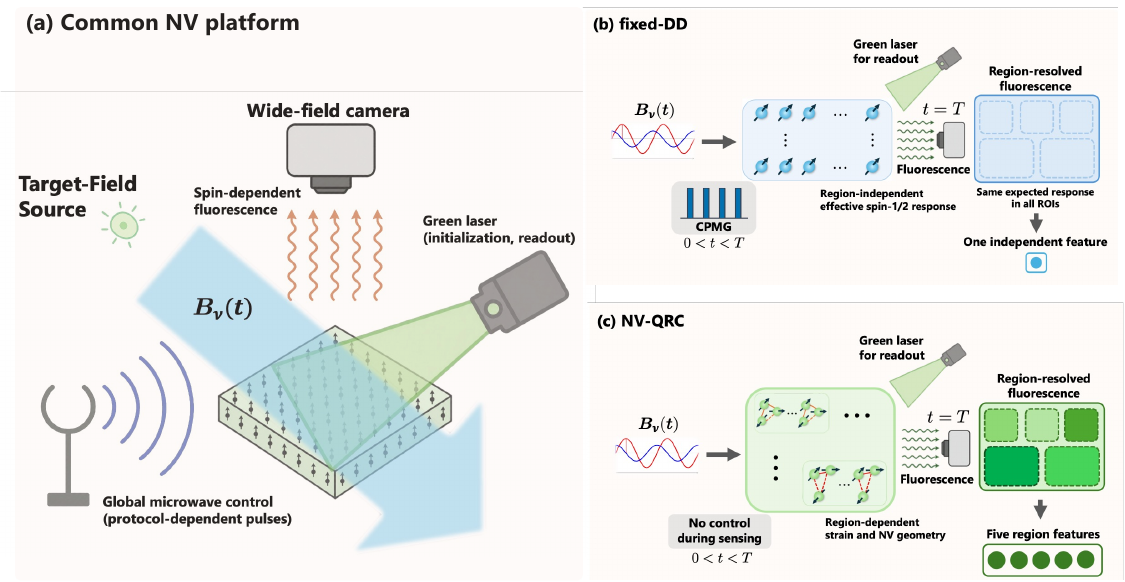}
\caption{Common measurement platform based on an NV ensemble and comparison of measurement protocols. (a) Wide-field platform assumed for both fixed-DD and NV-QRC. During the sensing interval, the NV ensemble is exposed to the target field $B_\nu(t)$. The same microwave operations are applied to all readout regions, and green-light initialization and readout are performed for all regions. At the end of sensing, a wide-field camera spatially resolves the spin-dependent fluorescence. (b) In the fixed-DD protocol considered in the main text, a CPMG sequence common to all regions and a region-independent response from an effective spin-1/2 system are assumed. Even when the fluorescence image is spatially resolved, the expected fluorescence response is the same in every region, so there is only one independent feature. (c) In NV-QRC, microwave control is not applied during interaction with the field. Instead, region-dependent natural evolution arising from transverse strain, NV configurations, and NV--NV dipolar interactions in the interaction-on model generates distinct spin responses that are read out as fluorescence from multiple regions. The illustration shows an example with five readout regions.}
\label{fig:nv_platform_protocol_overview}
\end{figure*}

We consider a situation in which the magnetic field associated with each event is localized to a region on the scale of several hundreds of nm. Measuring such a localized field requires placing the sensor near the target and acquiring field information on a spatial scale of at most several hundred nm. It is difficult for a macroscopic classical magnetic-field sensor to satisfy both the spatial-resolution requirement assumed in this study and the requirement to measure non-repeatable events at room temperature. We therefore use an NV center, which can access local fields near a surface with high spatial resolution, as the sensor~\cite{taylor2008highSensitivity}. In particular, we consider placing multiple readout regions within the area affected by the localized field and simultaneously acquiring region-resolved readout signals from a high-density NV ensemble containing sufficiently many NV centers in each region~\cite{pham2011magneticImaging}.

We assume ensemble readout of the NV centers and treat each readout signal as the ensemble expectation value of the corresponding observable. Accordingly, the numerical model does not include projection noise due to a finite number of NV centers or finite-shot noise in fluorescence detection.
The measurements considered here separate green-light initialization, interaction with the target field, and fluorescence readout in time. During interaction with the target field, the protocol may or may not include microwave control. We do not consider cw-ODMR-like measurement schemes in which green light and microwaves are applied continuously at the same time.
In experiments, optical initialization and fluorescence readout require finite times~\cite{degen2017quantum,barry2020sensitivity}. For short-duration events for which these times are comparable to
the initialization and readout time, it can be difficult to execute multiple measurement cycles sequentially under different conditions and obtain additional measurements from the same event. This is one experimental motivation for considering a setting in which the same event cannot be remeasured under different conditions.

The common wide-field measurement platform based on an NV ensemble is shown in Fig.~\ref{fig:nv_platform_protocol_overview}(a). At the end of sensing, a wide-field camera spatially resolves the fluorescence into region-specific readout signals. The next section defines the measurement protocols used on this common platform.
\subsection{Binary phase classification}
\label{sec:binary_phase_task}

As an analytically tractable benchmark, we consider binary phase classification of a sinusoidal field with task frequency $f_{\task}$, which is common and constant within one event group but unknown at the time of measurement. Let the class label be the random variable $Y\in\{0,1\}$ and the label realization of event $\nu$ be $j_\nu \in\{0,1\}$. We take the class priors to be
\begin{equation}
\Pr(Y=0)=\Pr(Y=1)=\frac12
\label{eq:equal_priors}
\end{equation}
Let the task frequency be $f_{\task}$, the angular frequency $\omega:=2\pi f_{\task}$, the sensing time be $T$, and the field scale be $s$. Let the nonnegative random variable $\Lambda\ge0$ represent the dimensionless random amplitude that differs between events, and write its realization for event $\nu$ as $\lambda_\nu$. The input field is
\begin{equation}
B_\nu(t)
:=
s\lambda_\nu\sin(\omega t+\phi_{j_\nu}),
\qquad 0\le t\le T
\label{eq:input_event}
\end{equation}
The waveform with fixed phase and amplitude is written as
\begin{equation}
B_{\phi,\lambda}(t)
:=
s\lambda\sin(\omega t+\phi)
\label{eq:input_field}
\end{equation}
and the two classes are
\begin{equation}
\phi_0:=0,
\qquad
\phi_1:=\frac{\pi}{2}
\label{eq:phase_classes}
\end{equation}
Thus, labels 0 and 1 correspond to the sine and cosine phases, respectively.

The random amplitude $\Lambda$ is independent of $Y$ and follows the same distribution in both classes. In the numerical calculations,
\begin{equation}
\begin{aligned}
\Lambda
&\sim\operatorname{LogNormal}(\mu_{\log},\sigma_{\log}^2),\\
\sigma_{\log}&=0.5,\\
\mu_{\log}&=-\frac{\sigma_{\log}^2}{2}=-0.125
\end{aligned}
\label{eq:lambda_lognormal}
\end{equation}
and $\mathbb E[\Lambda]=1$.

The value of $f_{\task}$ is unknown when the measurement protocol is fixed. In the numerical calculations, the measurement protocol is held fixed across all evaluated values of $f_{\task}$. Training and evaluation are performed separately for each $f_{\task}$: a classifier is trained on features of training events generated at that frequency and applied to test events at the same frequency. Each classifier receives only the features obtained from the measurement; $f_{\task}$ and $\lambda_\nu$ are not supplied as additional input variables.

\subsection{Feature maps and classification limits}

Let $\mathcal P$ denote a fixed measurement protocol. For the waveform $B_{\phi,\lambda}$ in Eq.~\eqref{eq:input_field}, let the $m_{\mathcal P}$-dimensional feature vector generated by the protocol at time $T$ be
\begin{equation}
\bm x_{\mathcal P}(\phi,\lambda)
\in\mathbb R^{m_{\mathcal P}}
\label{eq:physical_feature_map}
\end{equation}
Here $(\phi,\lambda)$ in parentheses denotes the phase and random-amplitude realization that specify the input waveform. Define the random variable representing the feature obtained from a random input event and its class-conditional distribution by
\begin{align}
\bm X_{\mathcal P}
&:=\bm x_{\mathcal P}(\phi_Y,\Lambda),
\label{eq:feature_random_variable}
\\
\mu_j^{\mathcal P}
&:=\Law(\bm X_{\mathcal P}\mid Y=j),
\qquad j=0,1
\label{eq:class_conditional_feature_law}
\end{align}
where $\Law$ denotes a probability distribution.

We apply the binary-classification limit introduced in Sec.~\ref{sec:background_classification} to the class-conditional feature distributions $\mu_0^{\mathcal P}$ and $\mu_1^{\mathcal P}$. Under the equal priors in Eq.~\eqref{eq:equal_priors}, define the Bayes-optimal accuracy based on the fixed features as
\begin{equation}
\Acc_{\mathcal P}^{\ast}
:=
\Acc^\ast(\mu_0^{\mathcal P},\mu_1^{\mathcal P})
\label{eq:protocol_bayes_accuracy}
\end{equation}
For a fixed protocol $\mathcal P$, we call a parameter point satisfying $\mu_0^{\mathcal P}=\mu_1^{\mathcal P}$ an indistinguishable point. At an indistinguishable point, no classifier using only the features in Eq.~\eqref{eq:feature_random_variable} can achieve a population accuracy greater than $1/2$, and $\Acc_{\mathcal P}^{\ast}=1/2$. This is a population limit distinct from empirical accuracy on a finite test set.

\section{Measurement protocols}
\label{sec:protocols}
Figure~\ref{fig:nv_platform_protocol_overview} schematically shows the fixed-DD and NV-QRC measurement protocols on the common wide-field platform based on an NV ensemble. The protocols differ in the control applied during sensing and in how they use spatially resolved fluorescence.

\subsection{Fixed-DD}
\label{sec:fixed_dd_protocol}

We call a protocol fixed-DD when the DD pulse sequence, state preparation, operation before readout, and readout method are specified before measuring the target events and are not changed based on the task frequency $f_{\task}$, class label, or measurement result of a target event. In the main text, we consider the case in which the same CPMG sequence is applied to all regions, all events, and all task frequencies used for evaluation, and one feature is obtained from the readout at time $T$. We use this protocol to analyze what information a frequency-selective one-dimensional feature map retains for unknown $f_{\task}$. Choosing the CPMG sequence according to $f_{\task}$ is not included in the problem setting defined in Sec.~\ref{sec:problem}.

We use the CPMG sequence defined by Eqs.~\eqref{eq:cpmg_pulse_times}--\eqref{eq:cpmg_toggling} in Sec.~\ref{sec:background_dd}. Its filter integral for the input waveform is
\begin{equation}
\Phi_{N_p}(\phi,\lambda)
:=
\int_0^T\eta_{N_p}(t)B_{\phi,\lambda}(t)\,dt.
\label{eq:dd_filter_integral}
\end{equation}

Set $\xi:=\omega T$. For the input $B_{\phi_j,\lambda}$, we call
$\alpha_j(N_p,\xi)$ the fixed-DD phase coefficient and define it by
\begin{equation}
\alpha_j(N_p,\xi)
:=
\gamma_e sT
\int_0^1
\eta_{N_p}(Tu)
\sin(\xi u+\phi_j)\,du.
\label{eq:alpha_cpmg}
\end{equation}
For fixed-DD, we consider an effective spin-1/2 system formed by the two levels used for the measurement rather than the full spin-1 triplet of the NV center. Let the basis of the effective two-level system be $\{\ket{\uparrow},\ket{\downarrow}\}$ and define
\begin{equation}
\ket{+}:=\frac{\ket{\uparrow}+\ket{\downarrow}}{\sqrt2},
\qquad
\sigma_y:=-\mathrm i\ket{\uparrow}\bra{\downarrow}
+\mathrm i\ket{\downarrow}\bra{\uparrow}
\label{eq:effective_qubit_readout_defs}
\end{equation}
We assume that sufficiently many independent copies of the effective spin-1/2 system prepared in $\ket{+}$ exist, that they are exposed to the same target event, and that they undergo the same CPMG sequence. A single ensemble readout of $\sigma_y$ at time $T$ is modeled as a noiseless estimate of its expectation value. The one-dimensional feature is therefore
\begin{equation}
x_{\DD}(\phi_j,\lambda)
:=
\sin[\alpha_j(N_p,\xi)\lambda]
\label{eq:dd_feature_sin}
\end{equation}
The feature random variable conditioned on class $j$ is
\begin{equation}
X_{\DD,j}
:=
\sin[\alpha_j(N_p,\xi)\Lambda],
\qquad j=0,1
\label{eq:dd_feature_random}
\end{equation}
Let the class-conditional distribution be $\mu_j^{\DD}:=\Law(X_{\DD,j})$ and define the Bayes-optimal accuracy based on $x_{\DD}$ by
\begin{equation}
\Acc_{\DD}^{\ast}
:=
\Acc^\ast(\mu_0^{\DD},\mu_1^{\DD})
\label{eq:dd_bayes_accuracy}
\end{equation}
The derivation and details are given in Appendices~\ref{app:dd_readout} and~\ref{app:dd_cpmg_sin}. An extension in which different CPMG sequences are applied in parallel to NV groups with different transition frequencies to obtain multiple features from the same event is discussed in Appendix~\ref{app:ideal_multi_dd}.

\subsection{NV-QRC}
\label{sec:nv_qrc_protocol}

We use a high-density NV ensemble as a physical reservoir that produces multiple region-specific features from a finite-duration magnetic-field waveform. We call this measurement protocol NV-QRC\@. Below, we write the component of the external field $\bm B_{\mathrm{ext}}$ introduced in Sec.~\ref{sec:background_nv} along the common NV axis as the scalar $B(t)$. During $0\le t\le T$, the NV centers are exposed to the external field $B(t)$, and the field information is encoded in the Hamiltonian through the Zeeman term. Region-dependent natural quantum evolution transforms this information before readout, and the fluorescence measured in each region at time $T$ provides the corresponding feature.

The measurement uses $\mR$ spatial readout regions $\mathcal R^{(r_1)},\ldots,\mathcal R^{(r_{\mR})}$ fixed in advance. State preparation, the external field, the sensing time, the operation before readout, and fluorescence acquisition are common to all regions, and no region-specific control sequence is assigned. The selected region set $\mathcal C:=\{r_1,\ldots,r_{\mR}\}$ is also fixed as part of the measurement protocol. Differences in the spatial configurations and transverse strains of the NV sites, as well as in interactions between NV centers, can produce different responses to the same waveform.

For a fixed readout region $r\in\mathcal C$, consider one $n$-spin system describing its local response and index the NV sites in the system by $i=1,\ldots,n$. Let $S_{a,i}^{(r)}$ ($a=x,y,z$) be the spin-1 operator acting on site $i$, and write $\bm S_i^{(r)}:=(S_{x,i}^{(r)},S_{y,i}^{(r)},S_{z,i}^{(r)})$. Let the site position be $\bm r_i^{(r)}$, and define the displacement between two sites and its unit vector by
\begin{equation}
\bm r_{ij}^{(r)}:=\bm r_i^{(r)}-\bm r_j^{(r)},
\qquad
\hat{\bm r}_{ij}^{(r)}:=\frac{\bm r_{ij}^{(r)}}{|\bm r_{ij}^{(r)}|}
\label{eq:dipole_position_defs}
\end{equation}
Define the transverse-strain operators by
\begin{align}
Q_{x,i}^{(r)}&:=(S_{x,i}^{(r)})^2-(S_{y,i}^{(r)})^2,
\\
Q_{y,i}^{(r)}&:=S_{x,i}^{(r)}S_{y,i}^{(r)}+S_{y,i}^{(r)}S_{x,i}^{(r)}
\end{align}
The Hamiltonian during the sensing interval is
\begin{equation}
H^{(r)}(t;B)
:=
H_D^{(r)}+H_E^{(r)}+H_Z^{(r)}(t;B)+H_{\mathrm{dip}}^{(r)}
\label{eq:nv_hamiltonian}
\end{equation}
with terms
\begin{align}
H_D^{(r)}
&:=D_{\mathrm{gs}}\sum_{i=1}^{n}(S_{z,i}^{(r)})^2,
\\
H_E^{(r)}
&:=\sum_{i=1}^{n}
\left(E_{x,i}^{(r)}Q_{x,i}^{(r)}+E_{y,i}^{(r)}Q_{y,i}^{(r)}\right),
\\
H_Z^{(r)}(t;B)
&:=\gamma_e B(t)\sum_{i=1}^{n}S_{z,i}^{(r)}
\label{eq:nv_local_terms}
\end{align}
Here $H_D^{(r)}$ represents the zero-field splitting, $H_E^{(r)}$ the region- and site-dependent transverse strain, and $H_Z^{(r)}(t;B)$ the Zeeman interaction due to the external field along the common NV axis. $D_{\mathrm{gs}}$ is the ground-state zero-field splitting, $E_{x,i}^{(r)}$ and $E_{y,i}^{(r)}$ are transverse-strain coefficients, and $\gamma_e$ is the electron-spin gyromagnetic ratio. The dipolar interaction is
\begin{equation}
\begin{split}
H_{\mathrm{dip}}^{(r)}
:={}&
\sum_{i<j}\frac{J_0}{|\bm r_{ij}^{(r)}|^3}
\bigl[
\bm S_i^{(r)}\!\cdot\!\bm S_j^{(r)}
\\[-2pt]
&-3(\bm S_i^{(r)}\!\cdot\!\hat{\bm r}_{ij}^{(r)})
(\bm S_j^{(r)}\!\cdot\!\hat{\bm r}_{ij}^{(r)})
\bigr]
\end{split}
\label{eq:dipole}
\end{equation}
Here $H_{\mathrm{dip}}^{(r)}$ is the magnetic dipolar interaction between NV sites in the same region, $J_0$ is its coefficient, and the sum runs over site pairs with $i<j$. We set $\hbar=1$ below and express the Hamiltonian in angular-frequency units. Numerical values of the coefficients, the generation of positions and transverse strains, and the spin-1 matrix representation are given in Sec.~\ref{sec:numerical_setup} and Appendix~\ref{app:hamiltonian_config}. In the NV-QRC considered here, we use the natural evolution generated by the Hamiltonian in Eq.~\eqref{eq:nv_hamiltonian} during the sensing interval after state preparation and do not include additional DD pulses or transverse microwave driving in the protocol.

After optically initializing each NV center in $\ket{0}$, define
\begin{equation}
\ket{B}:=\frac{\ket{+1}+\ket{-1}}{\sqrt2},
\qquad
\ket{D}:=\frac{\ket{+1}-\ket{-1}}{\sqrt2}
\label{eq:prep_basis_states}
\end{equation}
We then apply
\begin{equation}
U_{\mathrm{prep}}
:=
\ket{B}\bra{0}+\ket{0}\bra{B}+\ket{D}\bra{D}
\label{eq:uprep}
\end{equation}
to all sites to prepare $\ket{\psi_0}=\ket{B}^{\otimes n}$. The state at time $T$ in region $r$ exposed to input event $B_\nu$ is
\begin{equation}
\ket{\widetilde\psi_\nu^{(r)}(T)}
:=
\mathcal T\exp\!\left[-\mathrm i\int_0^T H^{(r)}(t;B_\nu)\,dt\right]
\ket{\psi_0}
\label{eq:qrc_sensing_state}
\end{equation}
where $\mathcal T$ is the time-ordering operator. Before readout, we apply the same $(U_{\mathrm{prep}}^{\otimes n})^\dagger$ to all regions to obtain
\begin{equation}
\ket{\psi_\nu^{(r)}}
:=
(U_{\mathrm{prep}}^{\otimes n})^\dagger
\ket{\widetilde\psi_\nu^{(r)}(T)}
\label{eq:qrc_output_state}
\end{equation}
The site-averaged readout observable for one $n$-spin system is
\begin{equation}
O_{\Reservoir}^{(r)}
:=
\frac1n\sum_{i=1}^{n}(S_{z,i}^{(r)})^2
\label{eq:qrc_readout_observable}
\end{equation}
and the region feature is defined as
\begin{equation}
x_{\Reservoir}^{(r)}[\nu]
:=
\bra{\psi_\nu^{(r)}}O_{\Reservoir}^{(r)}\ket{\psi_\nu^{(r)}}
\label{eq:qrc_component}
\end{equation}
The feature vector obtained from the fixed region set $\mathcal C$ is
\begin{equation}
\bm x_{\Reservoir}^{(\mathcal C)}[\nu]
:=
\bigl(x_{\Reservoir}^{(r_1)}[\nu],\ldots,
x_{\Reservoir}^{(r_{\mR})}[\nu]\bigr)^{\mathsf T}
\in\mathbb R^{\mR}
\label{eq:qrc_feature}
\end{equation}
For a prospective experimental implementation, we assume that each readout region contains many interacting NV centers and that fluorescence from the region is measured. In the numerical calculation, we instead assume that each readout region contains sufficiently many independent copies of an $n$-spin system composed of spin-1 particles, all evolving under the same region-specific Hamiltonian. Interactions between NV sites through $H_{\mathrm{dip}}^{(r)}$ may be present within each copy, whereas interactions between copies are neglected. A simultaneous fluorescence measurement of these copies is modeled as a noiseless estimate of the expectation value in Eq.~\eqref{eq:qrc_component}. Details are given in Sec.~\ref{sec:numerical_setup} and Appendix~\ref{app:hamiltonian_config}.

We call the model including $H_{\mathrm{dip}}^{(r)}$ in Eq.~\eqref{eq:dipole} the interaction-on model. The model that keeps the same NV center configurations, transverse strains, state preparation, external field, sensing time, and readout but sets $H_{\mathrm{dip}}^{(r)}=0$ is called the interaction-off model. As discussed in Sec.~\ref{sec:background_nv}, increasing the NV concentration to ensure sufficient numbers of NV centers and fluorescence photons in each region while retaining spatial resolution shortens the mean NV--NV distance, making the dipolar interaction difficult to neglect. We therefore use the interaction-on model as a physically motivated model for a high-density NV ensemble and the interaction-off model as a comparison baseline that isolates the contribution of dipolar interactions to the features and classification performance. Below, $x_{\Reservoir,\mathrm{off}}^{(r)}(\phi,\lambda)$ denotes the feature obtained from region $r$ when the waveform $B_{\phi,\lambda}$ is evaluated under the interaction-off model. The feature vector for the fixed region set $\mathcal C=\{r_1,\ldots,r_{\mR}\}$ is
\begin{equation}
\begin{aligned}
&
\bm{x}_{\Reservoir,\mathrm{off}}^{(\mathcal C)}(\phi,\lambda)
:=\\
&\quad
\left(
x_{\Reservoir,\mathrm{off}}^{(r_1)}(\phi,\lambda),\ldots,
x_{\Reservoir,\mathrm{off}}^{(r_{\mR})}(\phi,\lambda)
\right)^{\mathsf T}
\in\mathbb R^{\mR}.
\end{aligned}
\label{eq:qrc_off_parameter_feature}
\end{equation}

\section{Analytical results}
\label{sec:analytical_results}

\subsection{Coinciding feature distributions in fixed-DD}
\label{sec:fixed_dd_analytical}

\begin{proposition}[Sufficient condition for coinciding fixed-DD feature distributions]
\label{prop:dd_indistinguishable}
For the one-dimensional feature defined in Eq.~\eqref{eq:dd_feature_sin}, suppose that the random amplitude $\Lambda$ follows the same distribution in both classes and that
\begin{equation}
\alpha_0(N_p,\xi)=\alpha_1(N_p,\xi)
\label{eq:dd_alpha_equal}
\end{equation}
Then, for every amplitude realization $\lambda$,
$x_{\DD}(\phi_0,\lambda)=x_{\DD}(\phi_1,\lambda)$.
Consequently, $\mu_0^{\DD}=\mu_1^{\DD}$, and under equal class priors,
\begin{equation}
\Acc_{\DD}^{\ast}=\frac12.
\end{equation}
\end{proposition}

Under Eq.~\eqref{eq:dd_alpha_equal}, the two features coincide for every amplitude realization, so Proposition~\ref{prop:dd_indistinguishable} does not depend on the particular form of the random-amplitude distribution.
\begin{proposition}[Sufficient conditions for equal fixed-DD phase coefficients]
\label{prop:dd_alpha_equal_condition}
For a positive task frequency $f_{\task}$, set
$\theta:=\xi/N_p=\pi f_{\task}/f_{\DD}$. For fixed $N_p$ and $\xi$, if $\cos(\theta/2)\ne0$ and either (i) or (ii) below holds, then Eq.~\eqref{eq:dd_alpha_equal} holds.
\par\noindent\textnormal{(i)}
\begin{equation}
\theta=4\pi\ell,
\qquad
\ell\in\mathbb Z_{\ge1}
\label{eq:dd_zero_response_condition_main}
\end{equation}
\noindent\textnormal{(ii)}
$\theta\notin4\pi\mathbb Z_{\ge1}$ and either (a) or (b) below holds.
\begin{equation}
\begin{aligned}
\text{(a)}\quad
(\cos\xi,\sin\xi)
&=
\left((-1)^{N_p},0\right),
\\
\text{(b)}\quad
(\cos\xi,\sin\xi)
&=
\left(0,(-1)^{N_p}\right).
\end{aligned}
\label{eq:dd_indist_points_main}
\end{equation}

\end{proposition}

By Proposition~\ref{prop:dd_indistinguishable}, the feature distributions coincide under either condition in Proposition~\ref{prop:dd_alpha_equal_condition}, and the Bayes-optimal accuracy under equal priors is $1/2$. Proofs of both propositions and the case distinction are given in Appendix~\ref{app:dd_cpmg_sin}.

\subsection{Coinciding features in the interaction-off model}
\label{sec:interaction_off_symmetry}

\begin{proposition}[Feature equality for the binary phase task]
\label{prop:qrc_off_symmetry_indistinguishable}
Consider the binary phase-classification task defined in Sec.~\ref{sec:binary_phase_task} under the interaction-off model, with the readout defined in Sec.~\ref{sec:nv_qrc_protocol}.
\begin{samepage}
When the condition
\begin{equation}
\omega T=\frac{3\pi}{2}\pmod{2\pi}
\label{eq:off_phase_condition}
\end{equation}
is satisfied, the following equality holds for every random-amplitude realization $\lambda$ and every readout region $r$:
\begin{equation}
x_{\Reservoir,\mathrm{off}}^{(r)}(\phi_1,\lambda)
=
x_{\Reservoir,\mathrm{off}}^{(r)}(\phi_0,\lambda).
\label{eq:off_component_equality}
\end{equation}
\end{samepage}
Therefore, for any fixed region set $\mathcal C$,
\begin{equation}
\bm{x}_{\Reservoir,\mathrm{off}}^{(\mathcal C)}(\phi_1,\lambda)
=
\bm{x}_{\Reservoir,\mathrm{off}}^{(\mathcal C)}(\phi_0,\lambda).
\label{eq:off_vector_equality}
\end{equation}
\end{proposition}

This feature equality is not limited to the task considered here; it holds for any real signal and its time-reversed, sign-inverted counterpart. A proof for the general case is given in Appendix~\ref{app:interaction_off_proof}. Because the random amplitude $\Lambda$ is independent of the label $Y$ and follows the same distribution in both classes, the class-conditional feature distributions also coincide. Under equal priors, the Bayes-optimal accuracy is $1/2$.

\section{Numerical setup}
\label{sec:numerical_setup}

\subsection{Input data and classifier}

In the numerical calculations, the sensing time and field scale are
\begin{equation}
T=10~\mu\mathrm{s},
\qquad
s=1~\mu\mathrm{T}
\label{eq:numerical_time_field_scale}
\end{equation}
For each task frequency $f_{\task}$, we generate 400 training events and 100 test events using random amplitudes distributed according to Eq.~\eqref{eq:lambda_lognormal}. Labels are available for the training events, whereas the labels of the test events are not used to train the classifier. The two labels are balanced in both the training and test data, and each random-amplitude realization is assigned once to each label. This pairing makes the random-amplitude distributions of the two classes identical even for finite datasets.

For both fixed-DD and NV-QRC, we standardize the features and then classify the labels with a downstream ridge classifier. The standardization mean and variance and the classifier coefficients are estimated only from the training data, and the fixed standardization and classifier are applied to the test data. We use scikit-learn's \texttt{StandardScaler} and \texttt{RidgeClassifier} with ridge regularization coefficient $\alpha_{\mathrm{reg}}=10^{-4}$~\cite{pedregosa2011scikit}.

We denote the interaction-on/off model by $q\in\{\mathrm{on},\mathrm{off}\}$. For model $q$, a fixed region set $\mathcal C$, and a task frequency $f_{\task}$, the test accuracy is denoted by $\Acc_{\mathrm{test}}^{q}(f_{\task};\mathcal C)$ and is calculated as the fraction of the 100 test events classified correctly. This differs from the population Bayes-optimal accuracy in Eq.~\eqref{eq:bayes_limit}.

Each $f_{\task}$ represents an event group sharing the same unknown frequency, and groups with different common frequencies are evaluated as separate conditional tasks. For each task, one classifier is trained on the training features in that group and applied to the test features from the same group. The physical operations and region set $\mathcal C$ of NV-QRC are held fixed across all task frequencies in the set considered below.

\subsection{NV-QRC physical parameters and time evolution}

In the numerical calculations, we set $n=3$ for the region-level $n$-spin system introduced in Sec.~\ref{sec:nv_qrc_protocol}.
The fabrication of three proximate NV centers and dipolar coupling between NV pairs have been demonstrated~\cite{haruyama2019tripleNV}, suggesting that ensembles of such three-NV units may be feasible.
The NV center positions in each three-spin system are generated randomly within a $30\times30\times5~\mathrm{nm}^3$ rectangular box, with the positions differing between regions and pairwise distances constrained to be at least $3~\mathrm{nm}$. The Hamiltonian parameters are
\begin{align}
\frac{D_{\mathrm{gs}}}{2\pi}&\approx2.87~\mathrm{GHz},
&
\frac{\gamma_e}{2\pi}&\approx28~\mathrm{GHz/T},
\notag\\
\frac{J_0}{2\pi}&\approx52~\mathrm{MHz\,nm^3}
\label{eq:numerical_nv_parameters}
\end{align}
We use standard NV ground-state values of $D_{\mathrm{gs}}$ and $\gamma_e$~\cite{rondin2014magnetometry,degen2017quantum}, and the electron-spin dipolar coefficient for $J_0$~\cite{davis2023probing}. The transverse-strain components $E_{x,i}^{(r)}/2\pi$ and $E_{y,i}^{(r)}/2\pi$ are generated independently from Lorentzian distributions centered at zero with a half-width at half-maximum of $0.73~\mathrm{MHz}$, with values differing between regions. In the numerical calculations, each component is clipped to $[-7.3,7.3]~\mathrm{MHz}$ to avoid extreme values. The distribution width is chosen based on the broadening scale reported for optically detected magnetic resonance in a high-density NV ensemble~\cite{matsuzakiOpticallyDetectedMagnetic2016}.

The time evolution in Eq.~\eqref{eq:qrc_sensing_state} is calculated by evaluating the Hamiltonian at the midpoint of each interval and using a time-ordered product with step size $0.005~\mu\mathrm{s}$. For the main results, we generate $\Ncand=30$ region samples according to the rule above and enumerate all $\binom{30}{5}$ region sets to evaluate every possible set of $\mR=5$ readout regions. Different sets can share regions and are therefore not independent statistical samples. The interaction-on and interaction-off models use the same positions and transverse strains, with only $H_{\mathrm{dip}}^{(r)}$ removed in the latter. Detailed matrix representations and generation conditions are given in Appendix~\ref{app:hamiltonian_config}.

\subsection{Task-frequency sets and evaluation metric}

For the comparison between fixed-DD and NV-QRC, we use two fixed-DD settings: $f_{\DD}=1.0~\mathrm{MHz}$ and $f_{\DD}=1.25~\mathrm{MHz}$. We use the following set of 10 task frequencies $f_{\task}$:
\begin{equation}
\begin{split}
\mathcal F:=\{&1,\ 1.25,\ 1.625,\ 2.375,\ 3,\\
&3.375,\ 3.75,\ 4,\ 4.625,\ 5\}~\mathrm{MHz}
\end{split}
\label{eq:tested_grid}
\end{equation}
This set includes task frequencies that do and do not satisfy the fixed-DD feature-distribution equality condition derived in Sec.~\ref{sec:fixed_dd_analytical}. For NV-QRC, we define the minimum test accuracy when the same five-region set $\mathcal C$ is used for every task frequency in $\mathcal F$ by
\begin{equation}
A_{\min}^{q}(\mathcal C;\mathcal F)
:=
\min_{f_{\task}\in\mathcal F}
\Acc_{\mathrm{test}}^{q}(f_{\task};\mathcal C)
\label{eq:amin_def}
\end{equation}
For all $\binom{30}{5}$ sets of five regions selected from the 30 numerically generated region samples, we calculate $A_{\min}^{q}(\mathcal C;\mathcal F)$ and compute the fraction of region sets whose value is at least a specified threshold.

\section{Numerical results}
\label{sec:numerical_results}

\subsection{Frequency dependence of fixed-DD accuracy}
\label{sec:fixed_dd_numerical}

\begin{figure*}[t]
  \centering
  \includegraphics[width=0.94\textwidth]{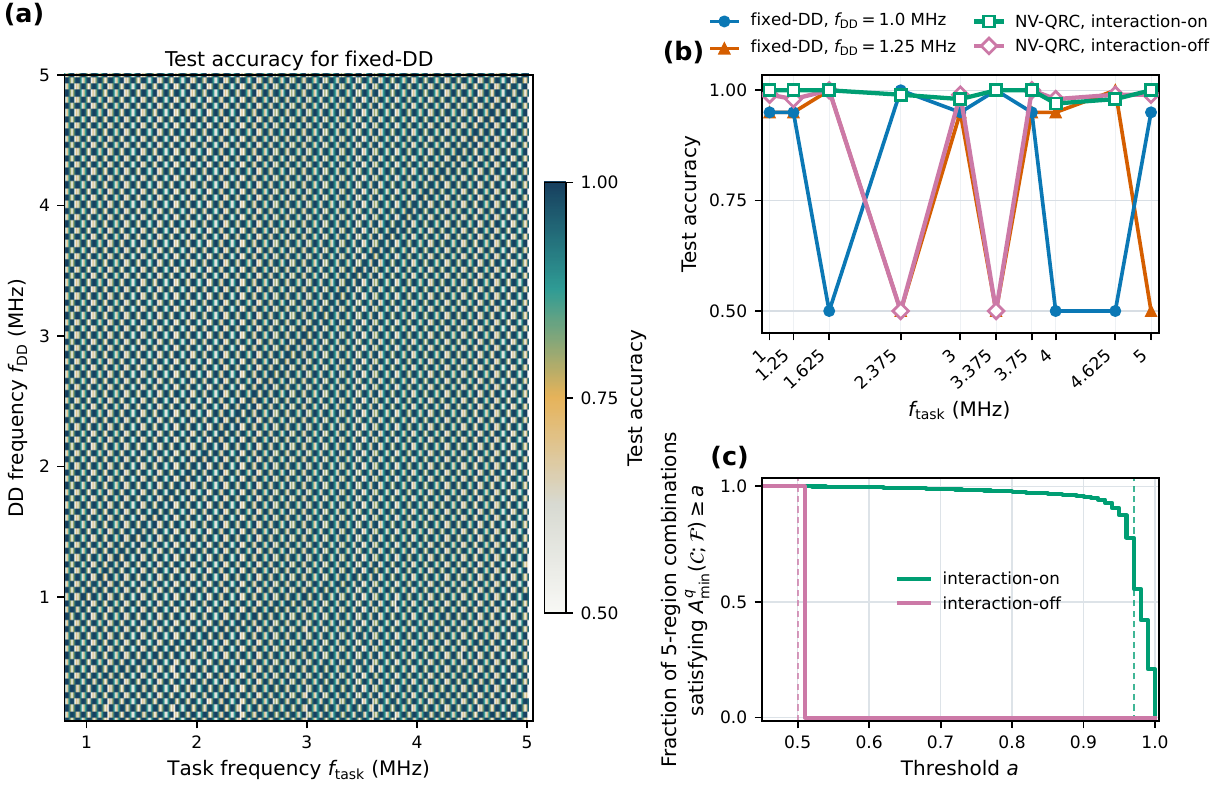}
  \caption{Comparison of accuracies for phase classification of non-repeatable events. (a) Test accuracy of fixed-DD obtained from the feature $x_{\DD}$ in Eq.~\eqref{eq:dd_feature_sin} at each grid point $(f_{\DD},f_{\task})$. (b) Test accuracies of fixed-DD with $f_{\DD}=1.0,1.25~\mathrm{MHz}$ and of the interaction-on/off NV-QRC models using the same five-region set $\mathcal C$ over the 10-point task-frequency set $\mathcal F$. The median of $A_{\min}^{\mathrm{on}}(\mathcal C;\mathcal F)$ over all $\binom{30}{5}$ sets is $0.97$, and the displayed $\mathcal C$ is one set attaining this value. (c) Fraction of all $\binom{30}{5}$ sets satisfying $A_{\min}^{q}(\mathcal C;\mathcal F)\ge a$. The sets may share regions and are therefore not mutually independent. Each dashed line indicates the median of $A_{\min}^{q}(\mathcal C;\mathcal F)$ over all sets.}
  \label{fig:main_results}
\end{figure*}

Figure~\ref{fig:main_results}(a) shows the test accuracy obtained by scanning the DD frequency $f_{\DD}$ and task frequency $f_{\task}$ for the fixed-DD protocol considered in the main text. We use $N_p=1,\ldots,100$, corresponding to $f_{\DD}=0.05,\ldots,5.0~\mathrm{MHz}$, and scan $f_{\task}=0.05,\ldots,5.0~\mathrm{MHz}$ in steps of $0.005~\mathrm{MHz}$. The color at each grid point $(f_{\DD},f_{\task})$ represents the accuracy for classifying test events at task frequency $f_{\task}$ using the CPMG sequence with the number of pulses $N_p$ that gives the specified $f_{\DD}$. The classifier receives only the feature $x_{\DD}$ in Eq.~\eqref{eq:dd_feature_sin}.

At $(f_{\DD},f_{\task})$ satisfying the sufficient condition for coinciding feature distributions shown in Sec.~\ref{sec:fixed_dd_analytical}, the test accuracy is $0.50$, in agreement with the analytical result. The indistinguishable task frequencies $f_{\task}$ depend on the DD frequency $f_{\DD}$.

\subsection{Comparison of fixed-DD and NV-QRC}
\label{sec:protocol_comparison_results}

Figure~\ref{fig:main_results}(b) compares the two fixed-DD settings, $f_{\DD}=1.0~\mathrm{MHz}$ and $1.25~\mathrm{MHz}$, with the interaction-on and interaction-off NV-QRC models over $\mathcal F$. The case $f_{\DD}=1.0~\mathrm{MHz}$ corresponds to $N_p=20$; at $f_{\task}=1.625,4.0,4.625~\mathrm{MHz}$, the sufficient condition for coinciding feature distributions in Sec.~\ref{sec:fixed_dd_analytical} is satisfied, and the test accuracy is $0.50$. The case $f_{\DD}=1.25~\mathrm{MHz}$ corresponds to $N_p=25$; for the same reason, the test accuracy is $0.50$ at $f_{\task}=2.375,3.375,5.0~\mathrm{MHz}$.

For the NV-QRC curves, we use a set $\mathcal C$ whose $A_{\min}^{\mathrm{on}}(\mathcal C;\mathcal F)$ equals the median value $0.97$ over all $\binom{30}{5}$ sets of five regions selected from the 30 numerically generated region samples. The same $\mathcal C$ is used for the interaction-off model. The positions and transverse-strain values of these five regions are given in Appendix~\ref{app:median_visualization_regions}.

For this $\mathcal C$, $A_{\min}^{\mathrm{on}}(\mathcal C;\mathcal F)=0.97$ and $A_{\min}^{\mathrm{off}}(\mathcal C;\mathcal F)=0.50$. At $f_{\task}=2.375,3.375~\mathrm{MHz}$, the interaction-off test accuracy $\Acc_{\mathrm{test}}^{\mathrm{off}}(f_{\task};\mathcal C)$ is $0.50$ at both frequencies. The interaction-on accuracy $\Acc_{\mathrm{test}}^{\mathrm{on}}(f_{\task};\mathcal C)$ at the same two frequencies is $0.99$ and $1.00$, respectively.

\subsection{Evaluation over all readout-region sets}
\label{sec:finite_pool_results}

Figure~\ref{fig:main_results}(c) shows the fraction, among all $142506$ sets obtained by selecting $\mR=5$ regions from the $\Ncand=30$ numerically generated region samples, for which $A_{\min}^{q}(\mathcal C;\mathcal F)$ is at least the threshold on the horizontal axis. For the interaction-on model, the fractions of sets with $A_{\min}^{\mathrm{on}}(\mathcal C;\mathcal F)$ at least $0.90$, $0.95$, and $0.97$ are $95.55\%$, $87.51\%$, and $55.75\%$, respectively. For the interaction-off model, no set satisfies $A_{\min}^{\mathrm{off}}(\mathcal C;\mathcal F)\ge0.90$. Thus, for $95.55\%$ of the $142506$ sets in the interaction-on model, the test accuracy is at least $0.90$ at all 10 frequencies in $\mathcal F$.

\subsection{Visualization of class-dependent features}
\label{sec:feature_geometry_results}

\begin{figure*}[t]
  \centering
  \includegraphics[width=0.94\textwidth]{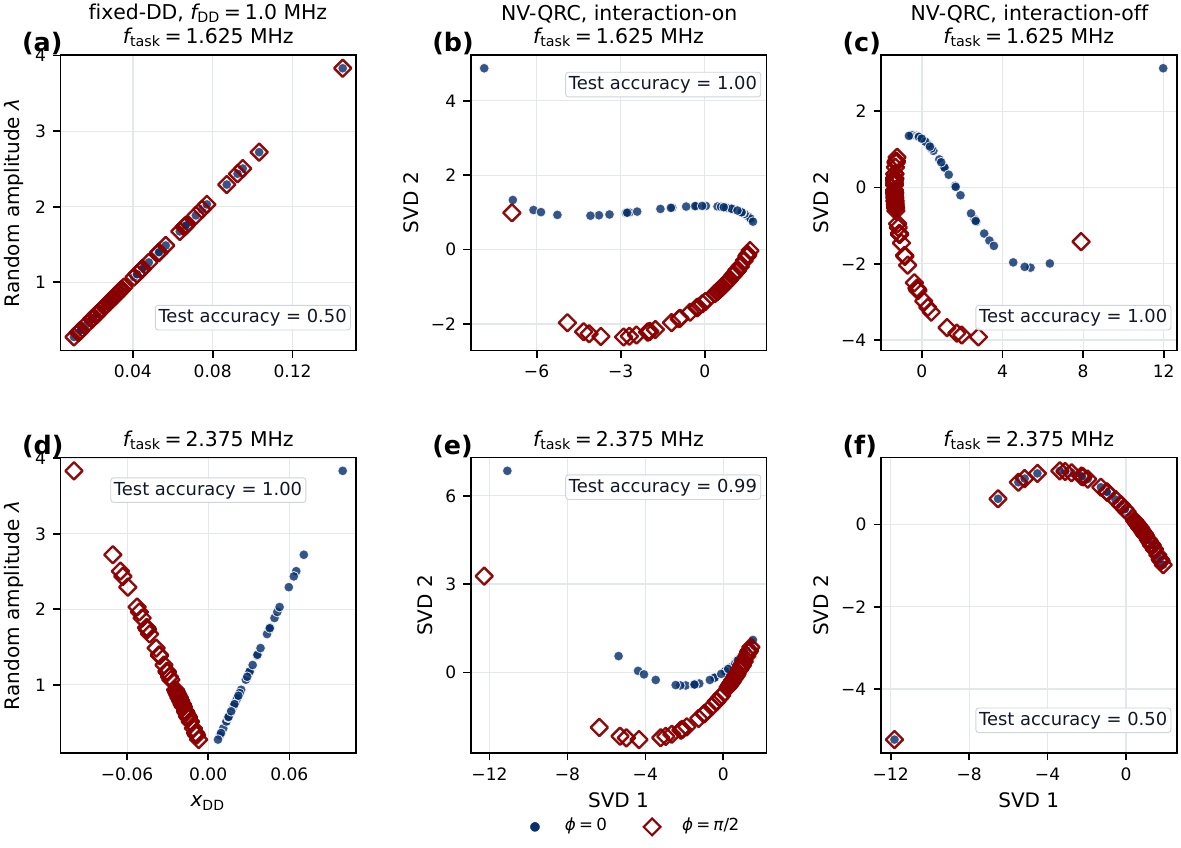}
  \caption{Class-dependent scatter plots of test events obtained from fixed-DD and NV-QRC\@. Panels (a)--(c) show $f_{\task}=1.625~\mathrm{MHz}$ and panels (d)--(f) show $2.375~\mathrm{MHz}$. Panels (a) and (d) show the fixed-DD feature $x_{\DD}$ obtained with $f_{\DD}=1.0~\mathrm{MHz}$ and the random-amplitude realization $\lambda$. Panels (b) and (e) show the projection of the five-dimensional feature vector of the interaction-on model onto the first two SVD axes, while panels (c) and (f) show the corresponding projection for the interaction-off model. Blue filled circles denote $\phi=0$, and red open diamonds denote $\phi=\pi/2$. NV-QRC uses the same five regions as in Fig.~\ref{fig:main_results}(b). The value in each panel is the test accuracy obtained by supplying $x_{\DD}$ for fixed-DD or the original five-dimensional feature vector for NV-QRC to the classifier. Neither $\lambda$ nor the two-dimensional SVD projection is supplied to the classifier.}
  \label{fig:feature_geometry}
\end{figure*}

Figure~\ref{fig:feature_geometry} shows the test events by class at $f_{\task}=1.625$ and $2.375~\mathrm{MHz}$. In panels (a) and (d) for fixed-DD, the horizontal axis is the feature $x_{\DD}$ supplied to the classifier and the vertical axis is the random-amplitude realization $\lambda$. The value of $\lambda$ is not supplied to the classifier. In panels (b), (c), (e), and (f) for NV-QRC, the five-dimensional feature vectors are standardized using the training-data mean and standard deviation, and the test data are projected onto the first two axes obtained by singular value decomposition (SVD) of the training data. The classifier is trained and evaluated on the original five-dimensional feature vectors rather than on this two-dimensional projection. Details are given in Appendix~\ref{app:feature_svd_projection}.

At $f_{\task}=1.625~\mathrm{MHz}$, the paired points from the two classes overlap for each $\lambda$ in fixed-DD panel (a), and the test accuracy is $0.50$. In contrast, the two class point clouds separate in the projection onto the first two SVD axes in panel (b) for the interaction-on model and panel (c) for the interaction-off model, and the test accuracy using the original five-dimensional feature vector is $1.00$ for both models.

At $f_{\task}=2.375~\mathrm{MHz}$, the two classes separate in fixed-DD panel (d), with a test accuracy of $1.00$. The two class point clouds also separate in panel (e) for the interaction-on model, with a test accuracy of $0.99$. In interaction-off panel (f), the paired points from the two classes overlap for each $\lambda$ and the test accuracy is $0.50$. This overlap agrees with the feature-vector equality in Eq.~\eqref{eq:off_vector_equality}.

\section{Discussion}
\label{sec:discussion}

Figure~\ref{fig:feature_geometry} shows that the two types of indistinguishable points considered in this work arise from different causes. At $f_{\task}=1.625~\mathrm{MHz}$, the fixed-DD feature distributions coincide, whereas both the interaction-on and interaction-off NV-QRC models discriminate the two classes. Thus, the class difference lost by fixed-DD at this frequency appears in the local responses of multiple regions with different transverse strains even without dipolar interactions. At $f_{\task}=2.375~\mathrm{MHz}$, in contrast, the time-reversal and sign-inversion symmetry of the interaction-off model maps the two inputs to the same feature in every region. Because this equality is independent of the choice and number of regions, adding noninteracting regions cannot enable discrimination between the two classes. At the same frequency, the interaction-on model evaluated here produces different features for the two classes, indicating that dipolar interactions remove the feature equality arising from the symmetry of the noninteracting dynamics.

In a high-density NV ensemble, region-dependent local responses and dipolar interactions within a region contribute to the features in different ways. Region-dependent responses map waveform differences that are absent from a single frequency-selective feature to distinct signals across multiple regions, whereas dipolar interactions preserve differences in the features that would otherwise be lost through the symmetry of the noninteracting dynamics. Interactions are increasingly being used as a resource in many-body quantum sensing~\cite{choi2020robustHamiltonian,zhouQuantumMetrologyStrongly2020,hughes2025stronglyInteracting,gao2026dressedState,montenegroReviewQuantumMetrology2025}, and the relation between information-processing performance and many-body dynamics has also been discussed in QRC~\cite{martinez-penaDynamicalPhaseTransitions2021,Kobayashi_2026}. Our results connect quantum sensing that uses many-body interactions as a measurement resource with QRC that uses many-body dynamics for information processing within a single setting: magnetic-field classification of non-repeatable events with a high-density NV ensemble.

The task of determining which member of a prescribed candidate set corresponds to the external signal acting on a quantum sensor has also been studied in the frameworks of quantum hypothesis testing and quantum state discrimination~\cite{tsangContinuousQuantumHypothesis2012,tsangNairFundamentalQuantumLimits2012,chaudhryDetectingPresenceWeak2015,liEnhancedQuantumHypothesis2025}. NV-QRC is related to quantum computational sensing because it uses the entire process from interaction with an external field to fluorescence readout to generate features and extract the information needed for classification~\cite{khanQuantumComputationalSensingAdvantage2025,khanQuantumComputationalSensing2026,prabhuQuantumComputationalDisplacement2026}.
From the perspective of QRC, task-dependent learning is delegated to a downstream classical learner, while the quantum reservoir itself remains fixed. It is therefore important for the reservoir to generate sufficiently informative features without being tailored to a specific task frequency. In our numerical study, the same fixed NV-QRC measurement protocol supported successful classification across the tested task frequencies, even though $f_{\task}$ was unknown at the time of measurement and task-specific training was confined to the downstream classifiers. This result suggests that QRC may be particularly useful for non-repeatable events when task-relevant signal conditions are not fully known at measurement time and the same event cannot later be remeasured under a different protocol.

Spatial multiplexing in QRC combines the outputs of multiple independent quantum reservoirs to enhance information-processing performance~\cite{nakajima2019boosting}. NV-QRC is analogous in that it combines distinct quantum responses from multiple spatial regions into a single feature vector. In NV-QRC, however, these responses are generated within a single high-density NV ensemble: fixed readout regions are exposed in parallel to the same magnetic-field event under common state preparation, sensing time, and readout operations. The response diversity arises from region-dependent transverse strains and NV-site configurations and, in the interaction-on model, from dipolar interactions within each region. Fluorescence from all regions is acquired simultaneously at the end of sensing, yielding multiple features without remeasuring the event at multiple times or under different measurement conditions. Thus, the NV dynamics and terminal fluorescence readout constitute the physical feature-generation process rather than post-processing applied to pre-acquired measurement data. We therefore compare fixed-DD and NV-QRC as measurement protocols for generating features from the same field waveform, using a common downstream classical classifier. The comparison concerns which information about the waveform each protocol retains in the measured features.

Let us discuss future perspectives of our research. We assume that each readout region contains many copies of a three-spin system evolving under the same region-specific Hamiltonian and use noiseless expectation values as features. In experiments, statistical fluctuations due to finite numbers of NV centers and detected photons, decoherence, and distributions of Hamiltonian parameters within a readout region affect the features. Because dissipation and quantum noise have also been reported to contribute to information processing in QRC~\cite{sanniaDissipationResourceQRC2024,kubotaTemporalInformationQuantumNoise2023,suzuki2022natural}, the effects of these processes on the features generated by the NV dynamics and on classification performance should be evaluated. In addition, NV electronic spins interact with spin baths consisting of $^{13}\mathrm{C}$ nuclear spins and substitutional-nitrogen defects known as P1 centers. The many-body dynamics of a P1 ensemble can appear in the decoherence response of an NV probe~\cite{davis2023probing}. In QRC, such non-Markovian dynamics have been reported to extend the retention of input history and improve performance on some time-series tasks~\cite{sanniaNonMarkovianityMemoryQRC2026,sasakiHamiltonianNonMarkovianQRC2025}. The influence of environmental spins should therefore not be treated only as decoherence; the effects of temporal correlations retained in the NV dynamics on features and classification performance should also be investigated.

We used binary phase classification of sine and cosine waves with random amplitudes as a benchmark for a theoretical analysis of the conditions under which feature distributions coincide. Future work should extend the task to more practical magnetic-field waveforms and examine models of high-density NV ensembles and measurement methods suited to the target signals and experimental conditions.

\section{Conclusion}
\label{sec:conclusion}

We proposed a method that uses quantum reservoir computing (QRC) for quantum sensing of non-repeatable events. A high-density NV center ensemble serves as both a quantum sensor and a physical reservoir: region-dependent spin responses generated by the input field and internal Hamiltonian are read out simultaneously as fluorescence from multiple regions, providing multiple features from one event.

For binary phase classification of sine and cosine waves with event-dependent random amplitudes, we used dynamical-decoupling magnetometry with one Carr--Purcell--Meiboom--Gill pulse sequence common to all events as the comparison method. In an effective two-level model with the same amplitude distribution for both classes, independent of the class label, the class-conditional feature distributions coincide under analytically derived conditions, so the classes cannot be discriminated from the single feature. Under the numerical conditions examined, the proposed model with dipolar interactions retained class-dependent differences in multiple features at the same frequencies. Even under conditions where the regional responses coincided in the model without dipolar interactions, this equality no longer held when the dipolar interactions were included.

These results demonstrate a role for QRC in quantum sensing of non-repeatable events: rather than using a quantum system for downstream processing of pre-acquired data, the method directly encodes the field information into the NV quantum state and uses the subsequent quantum dynamics and fluorescence readout to form classification features.

\clearpage
\appendix
\onecolumngrid
\begingroup
\section*{Notation summary}
\centering
\footnotesize
\begin{tabular}{p{0.24\textwidth}p{0.69\textwidth}}
\toprule
Notation & Meaning \\
\midrule
$Y,\ j_\nu$ & The binary label as a random variable and the realized label of event $\nu$.\\
$s,T,\omega$ & The magnetic-field amplitude scale, sensing time, and input angular frequency, with $\omega:=2\pi f_{\task}$.\\
$\phi_0,\phi_1$ & The two phases of the input waveform; in the main text, $\phi_0:=0$ and $\phi_1:=\pi/2$.\\
$\Lambda,\lambda_\nu,\mu_{\log},\sigma_{\log}$ & The random variable for the dimensionless event-dependent amplitude, its realization for event $\nu$, and the parameters of the normal distribution underlying the log-normal distribution. Numerical calculations use Eq.~\eqref{eq:lambda_lognormal}.\\
$f_{\task}, f_{\DD}$ & The task frequency, common to one event group but unknown at measurement time, and the DD frequency fixed before measurement, with $f_{\DD}:=N_p/(2T)$.\\
$N_p,\xi,\theta$ & The number of $\pi$ pulses in the CPMG sequence, $\xi:=\omega T$, and $\theta:=\xi/N_p=\pi f_{\task}/f_{\DD}$.\\
$\bm x_{\mathcal P}(\phi,\lambda),\bm X_{\mathcal P},m_{\mathcal P}$ & The feature vector generated by measurement protocol $\mathcal P$ from waveform $B_{\phi,\lambda}$, the random variable representing feature vectors from random input events, and the feature dimension.\\
$\eta_{N_p},\Phi_{N_p}$ & The CPMG toggling function and the corresponding filter integral.\\
$x_{\DD}, X_{\DD,j}$ & The feature obtained with fixed-DD and the feature random variable conditioned on class $j$.\\
$\alpha_j(N_p,\xi)$ & The phase coefficient for fixed-DD.\\
$\mu_j^{\mathcal P},\mu_j^{\DD}$ & The class-$j$-conditional distribution of the features generated by protocol $\mathcal P$ and its fixed-DD specialization.\\
$\mathcal R^{(r)}$ & An NV-QRC readout region; the superscript $(r)$ labels the region.\\
$n,\mR,\Ncand$ & The number of spin-1 NV sites in one $n$-spin system used to model the local response of region $r$, the number of readout regions, and the number of numerically generated readout-region samples. The main results use $n=3$, $\mR=5$, and $\Ncand=30$.\\
$\ket{B},\ket{D},U_{\mathrm{prep}}$ & The two single-spin states and the single-spin unitary used for NV-QRC state preparation, defined in Eqs.~\eqref{eq:prep_basis_states} and~\eqref{eq:uprep}.\\
$S_a,\ S_{a,i}^{(r)}$ & The single-NV spin-1 matrices and their embedded counterparts acting on the $i$th NV in region $r$ $(a=x,y,z)$.\\
$D_{\mathrm{gs}},\gamma_e,J_0$ & The NV ground-state zero-field splitting, the magnitude of the electron-spin gyromagnetic ratio, and the dipolar coefficient.\\
$E_{x,i}^{(r)},E_{y,i}^{(r)}$ & The transverse-strain components for each site and region.\\
$\ket{\psi_{\nu}^{(r)}}$ & The readout state obtained for event $\nu$ in region $r$ by applying $(U_{\mathrm{prep}}^{\otimes n})^\dagger$ to the state at time $T$.\\
$O_{\Reservoir}^{(r)}$ & The site-averaged readout observable for one $n$-spin system in region $r$, defined in Eq.~\eqref{eq:qrc_readout_observable}.\\
$x_{\Reservoir}^{(r)}[\nu],\bm{x}_{\Reservoir}^{(\mathcal C)}[\nu]$ & The feature component obtained from region $r$ at time $T$ for event $\nu$ and the feature vector obtained from a fixed region set $\mathcal C$. $x_{\Reservoir,\mathrm{off}}^{(r)}(\phi,\lambda)$ denotes the same feature evaluated for the interaction-off condition.\\
$\mathcal C$ & The set of readout-region labels fixed as part of the measurement protocol. In the main numerical evaluation, all sets satisfying $|\mathcal C|=\mR=5$ are enumerated from $\Ncand=30$ readout-region samples. These sets may share regions and are therefore not mutually independent. Appendix~\ref{app:qrc_m_dependence} also considers $\mR=1,\ldots,5$.\\
$q$ & The interaction-model label. $q\in\{\mathrm{on},\mathrm{off}\}$, where $\mathrm{on}$ denotes the interaction-on model including dipolar interactions and $\mathrm{off}$ denotes the interaction-off model obtained by removing only $H_{\mathrm{dip}}^{(r)}$.\\
$B^\sharp(t)$ & The waveform transformation $B^\sharp(t):=-B(T-t)$ used only in the general interaction-off proof in Appendix~\ref{app:interaction_off_proof}.\\
$\mathcal F$ & The set of 10 task frequencies $f_{\task}$ tested in the main text.\\
$\mathcal F_{\mathrm{fine}}$ & The set of 33 fine-grid task frequencies $f_{\task}$ used in Appendices~\ref{app:fine_frequency} and~\ref{app:qrc_m_dependence}.\\
$\Amin^{q}(\mathcal C;\mathcal F)$, $\Amin^{q}(\mathcal C;\mathcal F_{\mathrm{fine}})$ & For model $q$, the minimum test accuracy when $\mathcal C$ is fixed over all task frequencies in the 10-point main-text set or the 33-point fine grid used in Appendices~\ref{app:fine_frequency} and~\ref{app:qrc_m_dependence}, respectively.\\
\bottomrule
\end{tabular}
\par
\endgroup

\clearpage
\twocolumngrid
\section{Derivation of the fixed-DD feature}
\label{app:dd_readout}

We show, using an effective two-level model, that the feature $x_{\DD}$ used for fixed-DD in the main text is the expectation value of $\sigma_y$ at time $T$.
To distinguish the effective two-level basis used only in this section from the spin-1 basis of the NV center, we use arrow notation for its states:
\begin{equation}
\ket{\uparrow}
:=
\begin{pmatrix}
1\\
0
\end{pmatrix},
\quad
\ket{\downarrow}
:=
\begin{pmatrix}
0\\
1
\end{pmatrix},
\quad
\ket{+}
:=
\frac{\ket{\uparrow}+\ket{\downarrow}}{\sqrt2}.
\label{eq:effective_qubit_basis}
\end{equation}
The Pauli matrices in this basis are
\begin{align}
\sigma_z
&:=
\ket{\uparrow}\bra{\uparrow}
-
\ket{\downarrow}\bra{\downarrow}
=
\begin{pmatrix}
1 & 0\\
0 & -1
\end{pmatrix},
\notag\\
\sigma_y
&:=
-\mathrm i\ket{\uparrow}\bra{\downarrow}
+
\mathrm i\ket{\downarrow}\bra{\uparrow}
=
\begin{pmatrix}
0 & -\mathrm i\\
\mathrm i & 0
\end{pmatrix}.
\label{eq:effective_qubit_pauli}
\end{align}
The effective Hamiltonian induced by the magnetic field is
\begin{equation}
H_B(t)
:=
\frac{\gamma_e B(t)}{2}\sigma_z.
\end{equation}
We prepare the initial state in $\ket{+}$ and apply a CPMG sequence consisting of $N_p$ ideal $\pi$ pulses at the times specified by Eq.~\eqref{eq:cpmg_pulse_times} in the main text. The corresponding toggling function is $\eta_{N_p}(t)\in\{+1,-1\}$, defined in Eq.~\eqref{eq:cpmg_toggling}.
In the toggling frame, the Hamiltonian is
\begin{equation}
H_{\mathrm{togg}}(t)
:=
\frac{\gamma_e \eta_{N_p}(t)B(t)}{2}\sigma_z
\end{equation}
The time evolution up to time $T$ is
\begin{equation}
\begin{aligned}
U(T)
&:=
\exp\!\biggl[-\frac{\mathrm i}{2}\varphi(T)\sigma_z\biggr],\\
\varphi(T)
&:=
\gamma_e\int_0^T\eta_{N_p}(t)B(t)\,dt
\end{aligned}.
\end{equation}
The resulting state is
\begin{equation}
U(T)\ket{+}
=
\frac{
e^{-\mathrm i\varphi(T)/2}\ket{\uparrow}
+
e^{\mathrm i\varphi(T)/2}\ket{\downarrow}
}{\sqrt2}
\end{equation}
and hence
\begin{equation}
\bra{+}U^\dagger(T)\sigma_y U(T)\ket{+}
=
\sin\varphi(T).
\end{equation}
Thus, we take $\sin\varphi(T)$ as the feature $x_{\DD}$ in this work.
For the input waveform $B_{\phi,\lambda}(t)$, Eq.~\eqref{eq:dd_filter_integral} in the main text gives
\begin{equation}
\varphi(T)
=
\gamma_e\Phi_{N_p}(\phi,\lambda).
\end{equation}
Substituting this into $\bra{+}U^\dagger(T)\sigma_y U(T)\ket{+}=\sin\varphi(T)$ yields the feature $x_{\DD}$ defined in Eq.~\eqref{eq:dd_feature_sin} of the main text.

\section{Conditions for coincident fixed-DD feature distributions}
\label{app:dd_cpmg_sin}

This section derives the finite sum appearing in the main-text feature $x_{\DD}$ and classifies sufficient conditions under which the two phase classes are mapped to the same feature distribution.
Using the pulse times in the CPMG sequence, define
\begin{equation}
\xi:=\omega T,
\qquad
\theta
:=
\frac{\xi}{N_p}
=
\pi\frac{f_{\task}}{f_{\DD}}
\label{eq:theta_def}
\end{equation}
The boundary times, identical to those in Eq.~\eqref{eq:cpmg_toggling} of the main text, are
\begin{equation}
\begin{aligned}
t_0&:=0,
&t_{N_p+1}&:=T,\\
t_k&:=\left(k-\frac12\right)\frac{T}{N_p},
&k&=1,\ldots,N_p
\end{aligned}
\end{equation}
so that $\eta_{N_p}(t)=(-1)^k$ on $[t_k,t_{k+1})$.
For the input waveform $B_{\phi,\lambda}(t):=s\lambda\sin(\omega t+\phi)$, define
\begin{equation}
A_{N_p}(\phi)
:=
\Phi_{N_p}(\phi,1)
\end{equation}
By integrating over each interval, we obtain
\begin{align}
A_{N_p}(\phi)
&=
\frac{s}{\omega}
\sum_{k=0}^{N_p}
(-1)^k
\bigl[
\cos(\omega t_k+\phi)
\notag\\
&\hspace{5em}-
\cos(\omega t_{k+1}+\phi)
\bigr]
\notag\\
&=
\frac{s}{\omega}
\biggl\{
\cos\phi
\notag\\
&\quad+2\sum_{k=1}^{N_p}(-1)^k
\cos\!\biggl[\left(k-\frac12\right)\theta+\phi\biggr]
\notag\\
&\hspace{5em}
+(-1)^{N_p+1}\cos(\xi+\phi)
\biggr\}
\label{eq:cpmg_finite_sum}
\end{align}

We next evaluate the sum appearing in the second term of Eq.~\eqref{eq:cpmg_finite_sum}:
\[
\sum_{k=1}^{N_p}
(-1)^k
\cos\!\biggl[\left(k-\frac12\right)\theta+\phi\biggr].
\]
\begin{equation}
C_{N_p}
:=
\sum_{k=1}^{N_p}
(-1)^k
\exp\!\left[\mathrm i\left(k-\frac12\right)\theta\right]
\end{equation}
Then
\begin{align}
C_{N_p}
&=
e^{-\mathrm i\theta/2}
\sum_{k=1}^{N_p}
\left(-e^{\mathrm i\theta}\right)^k
\notag\\
&=
-
\frac{1-(-1)^{N_p}e^{\mathrm i\xi}}
{2\cos(\theta/2)}.
\label{eq:cpmg_complex_sum}
\end{align}
Here we assume $\cos(\theta/2)\ne0$.
Combining the first and third terms of Eq.~\eqref{eq:cpmg_finite_sum} gives
\begin{equation}
\begin{aligned}
&
\cos\phi+(-1)^{N_p+1}\cos(\xi+\phi)
\\
&\quad=
\operatorname{Re}
\left[
e^{\mathrm i\phi}
\left\{
1-(-1)^{N_p}e^{\mathrm i\xi}
\right\}
\right].
\end{aligned}
\end{equation}
The second term of Eq.~\eqref{eq:cpmg_finite_sum} is
\begin{equation}
\begin{aligned}
&
2\sum_{k=1}^{N_p}(-1)^k
\cos\!\biggl[\left(k-\frac12\right)\theta+\phi\biggr]
\\
&\quad=
2\operatorname{Re}\!\left(e^{\mathrm i\phi}C_{N_p}\right).
\end{aligned}
\end{equation}
Therefore, when $\cos(\theta/2)\ne0$,
\begin{equation}
\begin{split}
\gamma_e A_{N_p}(\phi)
=
g_{N_p}(\xi)
\bigl\{
&[1-(-1)^{N_p}\cos\xi]\cos\phi\\
&+(-1)^{N_p}\sin\xi\sin\phi
\bigr\}
\end{split}
\label{eq:cpmg_closed_form}
\end{equation}
where
\begin{equation}
\begin{aligned}
g_{N_p}(\xi)
&:=
\frac{\kappa_{\DD}}{\xi}
\biggl[
1-\sec\!\left(\frac{\theta}{2}\right)
\biggr],
\\
\kappa_{\DD}
&:=
\gamma_e sT.
\end{aligned}
\label{eq:cpmg_g_def}
\end{equation}

For the two phases in the main text, $\phi_0:=0$ and $\phi_1:=\pi/2$,
\begin{align}
\alpha_0(N_p,\xi)
&=
g_{N_p}(\xi)
\bigl[1-(-1)^{N_p}\cos\xi\bigr],
\label{eq:alpha0_closed}
\\
\alpha_1(N_p,\xi)
&=
g_{N_p}(\xi)
(-1)^{N_p}\sin\xi.
\label{eq:alpha1_closed}
\end{align}
For fixed $N_p$ and $\xi$, the features of the two classes are
\begin{equation}
X_{\DD,j}
=
\sin[\alpha_j(N_p,\xi)\Lambda],
\qquad
j=0,1.
\end{equation}
Because the random amplitude $\Lambda$ has the same distribution in both classes, the distributions of $X_{\DD,0}$ and $X_{\DD,1}$ coincide whenever $\alpha_0(N_p,\xi)=\alpha_1(N_p,\xi)$.

First consider the case $\cos(\theta/2)\ne0$ and $g_{N_p}(\xi)\ne0$. From Eqs.~\eqref{eq:alpha0_closed} and~\eqref{eq:alpha1_closed},
\begin{align}
\alpha_0=\alpha_1
&\Longleftrightarrow
1-(-1)^{N_p}\cos\xi
=
(-1)^{N_p}\sin\xi
\notag\\
&\Longleftrightarrow
\sin\xi+\cos\xi
=
(-1)^{N_p}.
\label{eq:dd_indist_condition}
\end{align}
Equivalently, either (a) or (b) holds:
\begin{equation}
\begin{aligned}
\text{(a)}\quad
\left(\cos\xi,\sin\xi\right)
&=
\left((-1)^{N_p},0\right),
\\
\text{(b)}\quad
\left(\cos\xi,\sin\xi\right)
&=
\left(0,(-1)^{N_p}\right).
\end{aligned}
\label{eq:dd_indist_points}
\end{equation}
In this case, the feature $x_{\DD}$ maps the two phase classes to the same one-dimensional distribution, and the Bayes-optimal accuracy is $1/2$.

Next consider the case $\cos(\theta/2)\ne0$ and $g_{N_p}(\xi)=0$. From Eq.~\eqref{eq:cpmg_g_def}, for positive task frequency $f_{\task}$, $g_{N_p}(\xi)=0$ when
\begin{equation}
\theta=4\pi\ell,
\qquad
\ell\in\mathbb Z_{\ge1}.
\label{eq:dd_zero_response_condition}
\end{equation}
Under this condition,
\begin{equation}
\alpha_0=\alpha_1=0,
\qquad
X_{\DD,0}=X_{\DD,1}=0.
\end{equation}
In this case, the features of both classes again coincide at zero.

Finally, when $\cos(\theta/2)=0$, Eqs.~\eqref{eq:cpmg_complex_sum} and~\eqref{eq:cpmg_closed_form} cannot be used, so we evaluate the finite sum in Eq.~\eqref{eq:cpmg_finite_sum} directly. This condition is
\begin{equation}
\theta=(2p+1)\pi,
\qquad
p\in\mathbb Z_{\ge0}
\label{eq:dd_resonance_theta}
\end{equation}
or, equivalently,
\begin{equation}
f_{\DD}
=
\frac{f_{\task}}{2p+1}
\label{eq:dd_resonance_frequency}
\end{equation}
In this case,
\begin{equation}
\alpha_0=0,
\qquad
\alpha_1
=
(-1)^p
\frac{2\kappa_{\DD}}{(2p+1)\pi}
\label{eq:dd_resonance_alpha}
\end{equation}
Therefore, Eq.~\eqref{eq:dd_feature_random} gives
\begin{equation}
\begin{aligned}
X_{\DD,0}&=0,\\
X_{\DD,1}
&=
\sin\!\biggl[
(-1)^p
\frac{2\kappa_{\DD}}{(2p+1)\pi}
\Lambda
\biggr].
\end{aligned}
\end{equation}
Under this resonance condition, the relation between the CPMG pulse times and the task frequency $f_{\task}$ causes only one phase class to have an amplitude-dependent feature. This is an analysis of the response at fixed $f_{\DD}$ and does not represent a procedure for choosing the CPMG sequence according to the unknown $f_{\task}$.

Thus, when condition (i) or (ii) of Proposition~\ref{prop:dd_alpha_equal_condition} holds, $\alpha_0(N_p,\xi)=\alpha_1(N_p,\xi)$. Moreover, Eq.~\eqref{eq:dd_feature_random} shows that the two features coincide for every amplitude realization, which proves Proposition~\ref{prop:dd_indistinguishable}.

\section{NV-QRC model and numerical conditions}
\label{app:hamiltonian_config}

The NV-QRC calculations in the main text use the Hamiltonian~\eqref{eq:nv_hamiltonian} for $n:=3$ spin-1 NV centers in each readout region. We assume that each region contains many independent copies of this $n$-spin system, all evolving under the same region-specific Hamiltonian, so that the fluorescence from the region is modeled by the expectation value in Eq.~\eqref{eq:qrc_component}.
The basis order for a single spin-1 NV center is
\begin{equation}
\ket{+1}
:=
\begin{pmatrix}
1\\
0\\
0
\end{pmatrix},
\quad
\ket{0}
:=
\begin{pmatrix}
0\\
1\\
0
\end{pmatrix},
\quad
\ket{-1}
:=
\begin{pmatrix}
0\\
0\\
1
\end{pmatrix}
\label{eq:spin1_basis}
\end{equation}
The states used for state preparation in the main text are
\begin{equation}
\ket{B}
:=
\frac{\ket{+1}+\ket{-1}}{\sqrt2},
\qquad
\ket{D}
:=
\frac{\ket{+1}-\ket{-1}}{\sqrt2}.
\end{equation}
The matrix representations below use the basis order $(\ket{+1},\ket{0},\ket{-1})$ in Eq.~\eqref{eq:spin1_basis}.
The dimensionless spin-1 matrices in this basis are
\begin{align}
S_x
&=
\ket{B}\bra{0}
+
\ket{0}\bra{B}
=
\frac{1}{\sqrt2}
\begin{pmatrix}
0 & 1 & 0\\
1 & 0 & 1\\
0 & 1 & 0
\end{pmatrix},
\notag\\
S_y
&=
-\mathrm i\ket{D}\bra{0}
+
\mathrm i\ket{0}\bra{D}
=
\frac{1}{\sqrt2}
\begin{pmatrix}
0 & -\mathrm i & 0\\
\mathrm i & 0 & -\mathrm i\\
0 & \mathrm i & 0
\end{pmatrix},
\notag\\
S_z
&=
\ket{B}\bra{D}
+
\ket{D}\bra{B}
=
\begin{pmatrix}
1 & 0 & 0\\
0 & 0 & 0\\
0 & 0 & -1
\end{pmatrix}.
\label{eq:spin1_matrices}
\end{align}
For each region, the NV center positions are generated independently and uniformly within a $30\times30\times5~\mathrm{nm}^3$ rectangular volume, subject to a minimum separation of $3~\mathrm{nm}$. For site $i$ in region $r$, define its position, the displacement between two sites, and the corresponding unit vector by
\begin{equation}
\begin{aligned}
\bm r_i^{(r)}
&:=
(x_i^{(r)},y_i^{(r)},z_i^{(r)}),\\
\bm r_{ij}^{(r)}
&:=
\bm r_i^{(r)}-\bm r_j^{(r)},\\
\hat{\bm r}_{ij}^{(r)}
&:=
\frac{\bm r_{ij}^{(r)}}{|\bm r_{ij}^{(r)}|}
\end{aligned}
\label{eq:supp_position_defs}
\end{equation}
We used the following values in the numerical calculations, although the main text reports the physical scales only approximately. The electron-spin gyromagnetic ratio and zero-field splitting are
\begin{equation}
\begin{aligned}
\frac{\gamma_e}{2\pi}
&:=
28024.95164~\mathrm{MHz/T},
\\
\frac{D_{\mathrm{gs}}}{2\pi}
&:=
2870~\mathrm{MHz}.
\end{aligned}
\end{equation}
The transverse-strain components $E_{x,i}^{(r)}/2\pi$ and $E_{y,i}^{(r)}/2\pi$ are generated independently for each site and region from Lorentzian distributions centered at $0$ with half-width at half-maximum $0.73~\mathrm{MHz}$. To avoid extreme values, samples outside $[-7.3,7.3]~\mathrm{MHz}$ are clipped to the nearest boundary.
This strain scale is based on the broadening scale reported in optically detected magnetic resonance experiments on dense NV ensembles, whereas the Lorentzian distribution, the independence across components and sites, and the clipping are model choices in this work~\cite{matsuzakiOpticallyDetectedMagnetic2016}.

The interaction-on model includes the dipolar interaction in Eq.~\eqref{eq:dipole}, with dipolar coefficient
\begin{equation}
\frac{J_0}{2\pi}
=
52.04101673450712~\mathrm{MHz\,nm^3}
\end{equation}
We take Davis et al.'s rounded value $J_0/(2\pi)=52~\mathrm{MHz\,nm^3}$ for the magnetic dipolar coefficient between electron spins as a reference and use the value above in the present numerical calculations~\cite{davis2023probing}.
The interaction-off model removes only $H_{\mathrm{dip}}^{(r)}$ and keeps all other Hamiltonian parameters and readout conditions unchanged.
Time evolution is calculated using a time-ordered product with the Hamiltonian evaluated at the midpoint of each interval and a time step of $0.005~\mu{\mathrm{s}}$. In the classifier implementation, a constant classifier returning the majority class is used when all selected features are numerically constant. Because the test labels are balanced in these data, its test accuracy is $1/2$.

\subsection{Parameters of the readout regions used for the main results}
\label{app:median_visualization_regions}

For the NV-QRC data in Fig.~\ref{fig:main_results}(b) and Fig.~\ref{fig:feature_geometry}, we use a region set $\mathcal C$ whose $\Amin^{\mathrm{on}}(\mathcal C;\mathcal F)$ equals the median, $0.97$, among all $\binom{30}{5}$ sets formed by choosing five regions from the 30 numerically generated samples. The same $\mathcal C$ is used for the interaction-off model.
The physical quantities defining these five regions are listed in Tables~\ref{tab:representative_region_sites} and~\ref{tab:representative_region_distances}.
The region labels $r=1,\ldots,5$ are local labels for the five regions in $\mathcal C$. For this $\mathcal C$, $\Amin^{\mathrm{on}}(\mathcal C;\mathcal F)=0.97$, equal to the median over all $\binom{30}{5}$ sets. The positions and transverse-strain values in Table~\ref{tab:representative_region_sites} identify the set $\mathcal C$ used in the figures.
The positions and strain values are the same for the interaction-on and interaction-off models; the latter removes only $H_{\mathrm{dip}}^{(r)}$ in Eq.~\eqref{eq:dipole}.
Table~\ref{tab:representative_region_distances} lists, in addition to the distances, the distance-dependent coefficient multiplying the spin-operator part of Eq.~\eqref{eq:dipole},
\[
\mathcal J_{ij}^{(r)}
:=
\frac{J_0}{2\pi|\bm r_{ij}^{(r)}|^3}
\]
in units of MHz.

\begin{table*}[t]
\centering
\scriptsize
\begin{tabular}{cccccc}
\toprule
Region $r$ & NV $i$ & $x_i^{(r)}$ (nm) & $y_i^{(r)}$ (nm) & $z_i^{(r)}$ (nm) & $(E_{x,i}^{(r)},E_{y,i}^{(r)})/2\pi$ (MHz) \\
\midrule
1 & 1 & 6.638 & 21.418 & 0.440 & $(-0.525,\ 0.352)$ \\
1 & 2 & 22.840 & 17.633 & 4.373 & $(0.604,\ -0.116)$ \\
1 & 3 & 17.311 & 21.705 & 0.256 & $(0.547,\ -0.534)$ \\
2 & 1 & 12.780 & 3.431 & 2.912 & $(1.453,\ -3.957)$ \\
2 & 2 & 5.959 & 26.438 & 4.775 & $(-0.607,\ 2.457)$ \\
2 & 3 & 14.503 & 12.720 & 0.567 & $(2.646,\ 0.075)$ \\
3 & 1 & 18.376 & 11.957 & 3.878 & $(-1.106,\ -0.681)$ \\
3 & 2 & 23.061 & 22.045 & 3.266 & $(7.300,\ 2.245)$ \\
3 & 3 & 17.609 & 24.803 & 3.111 & $(1.333,\ 0.135)$ \\
4 & 1 & 0.130 & 26.039 & 0.530 & $(-0.272,\ 0.295)$ \\
4 & 2 & 17.939 & 1.304 & 4.139 & $(-2.710,\ 1.785)$ \\
4 & 3 & 21.282 & 20.035 & 2.292 & $(0.703,\ 7.300)$ \\
5 & 1 & 11.882 & 9.366 & 4.640 & $(0.480,\ 2.351)$ \\
5 & 2 & 29.878 & 10.727 & 1.000 & $(7.300,\ 0.250)$ \\
5 & 3 & 3.311 & 17.407 & 2.490 & $(7.300,\ -7.300)$ \\
\bottomrule
\end{tabular}
\caption{Positions and transverse-strain values of the five readout regions used in Figs.~\ref{fig:main_results}(b) and~\ref{fig:feature_geometry}. Positions are given in nm within the rectangular sampling volume, and strain values are given in MHz.}
\label{tab:representative_region_sites}
\end{table*}

\begin{table*}[t]
\centering
\scriptsize
\begin{tabular}{ccccccc}
\toprule
Region $r$
& $|\bm r_{12}^{(r)}|$ (nm) & $\mathcal J_{12}^{(r)}$ (MHz)
& $|\bm r_{13}^{(r)}|$ (nm) & $\mathcal J_{13}^{(r)}$ (MHz)
& $|\bm r_{23}^{(r)}|$ (nm) & $\mathcal J_{23}^{(r)}$ (MHz) \\
\midrule
1 & 17.097 & 0.01041 & 10.678 & 0.04274 & 8.006 & 0.10140 \\
2 & 24.069 & 0.00373 & 9.734 & 0.05642 & 16.700 & 0.01117 \\
3 & 11.140 & 0.03765 & 12.892 & 0.02428 & 6.112 & 0.22788 \\
4 & 30.692 & 0.00180 & 22.058 & 0.00485 & 19.117 & 0.00745 \\
5 & 18.411 & 0.00834 & 11.948 & 0.03051 & 27.434 & 0.00252 \\
\bottomrule
\end{tabular}
\caption{NV--NV distances obtained from the positions in Table~\ref{tab:representative_region_sites} and the distance-dependent coefficient $\mathcal J_{ij}^{(r)}$ in Eq.~\eqref{eq:dipole}. The coefficients are calculated from the double-precision coordinates used in the computation; only the coordinates and distances in the table are rounded to the displayed digits. The directional dependence in Eq.~\eqref{eq:dipole} is determined by the coordinates in Table~\ref{tab:representative_region_sites}.}
\label{tab:representative_region_distances}
\end{table*}

\section{Two-dimensional projection of NV-QRC features}
\label{app:feature_svd_projection}

We define the two-dimensional coordinates used in the NV-QRC panels of Fig.~\ref{fig:feature_geometry}. Fix a model $q\in\{\mathrm{on},\mathrm{off}\}$, a task frequency $f_{\task}$, and a region set $\mathcal C$, and denote the numbers of training and test samples by $N_{\mathrm{train}}$ and $N_{\mathrm{test}}$, respectively. The feature vector from each event has dimension $\mR$. We standardize the training and test features using the means and standard deviations of the individual training-feature components, as described in Sec.~\ref{sec:background_rc}. Let the matrices whose rows are the standardized feature vectors be
\begin{equation}
\bm Z_{\mathrm{train}}\in\mathbb R^{N_{\mathrm{train}}\times\mR},
\qquad
\bm Z_{\mathrm{test}}\in\mathbb R^{N_{\mathrm{test}}\times\mR}
\label{eq:svd_standardized_feature_matrices}
\end{equation}
The test data are standardized using the same means and standard deviations obtained from the training data.

Let $\bm v_1,\bm v_2\in\mathbb R^{\mR}$ be the right singular vectors of $\bm Z_{\mathrm{train}}$ associated with its largest and second-largest singular values, respectively, and define
\begin{equation}
\bm V_2
:=
\bigl(\bm v_1\ \bm v_2\bigr)
\in\mathbb R^{\mR\times2}
\label{eq:svd_projection_axes}
\end{equation}
The two-dimensional coordinates of the test samples shown in the figure are
\begin{equation}
\bm Z_{\mathrm{test}}^{(2)}
:=
\bm Z_{\mathrm{test}}\bm V_2
\in\mathbb R^{N_{\mathrm{test}}\times2}
\label{eq:svd_test_projection}
\end{equation}
The rows of $\bm Z_{\mathrm{test}}^{(2)}$ are plotted in Fig.~\ref{fig:feature_geometry}. Thus, the means and standard deviations used for standardization, as well as the projection axes $\bm V_2$, are all determined from the training data alone. The figure uses $N_{\mathrm{train}}=400$, $N_{\mathrm{test}}=100$, and $\mR=5$. This two-dimensional projection is used only for visualization; the ridge classifier is trained and evaluated using the $\mR$-dimensional standardized feature vectors before projection.

\section{Fine frequency sweep and indistinguishable points in the interaction-off model}
\label{app:fine_frequency}

Figure~\ref{fig:main_results} uses the 10-point set of task frequencies $f_{\task}$ defined in Eq.~\eqref{eq:tested_grid}. Here we use
\begin{equation}
\mathcal F_{\mathrm{fine}}
:=
\{1.0+0.125\ell_f:\ \ell_f=0,\ldots,32\}~\mathrm{MHz}
\label{eq:fine_grid}
\end{equation}
We use the 33 points from $1.0$ to $5.0~\mathrm{MHz}$ defined above to test for performance degradation between the 10 main-grid frequencies. Following the setup in Sec.~\ref{sec:binary_phase_task}, each task frequency $f_{\task}$ is treated as a separate conditional task for an event group sharing the same unknown frequency, with 400 training samples and 100 test samples for each task. The physical protocol and region set $\mathcal C$ are held fixed across all 33 frequencies.

\begin{figure*}[t]
  \centering
  \includegraphics[width=0.98\linewidth]{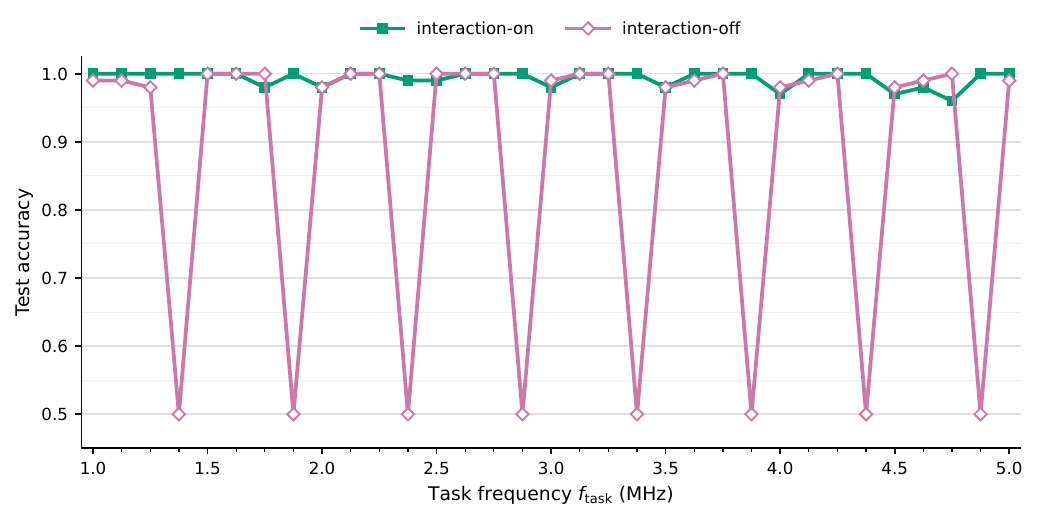}
  \caption{
  Fine frequency sweep for the five regions used in Fig.~\ref{fig:main_results}(b).
  The task frequency $f_{\task}$ takes the 33 values in Eq.~\eqref{eq:fine_grid}; the same five regions are used for the interaction-on and interaction-off models.
  For the interaction-off model, the test accuracy drops to $0.50$ at the 8 points $f_{\task}=1.375+0.5k~\mathrm{MHz}$, where $k=0,\ldots,7$.
  }
  \label{fig:fine_frequency_selected_subset}
\end{figure*}

The results for the same $\mathcal C$ as in Fig.~\ref{fig:main_results}(b) are shown in Fig.~\ref{fig:fine_frequency_selected_subset} and Table~\ref{tab:fine_frequency_selected_subset}.
For this $\mathcal C$, $\Amin^{\mathrm{on}}(\mathcal C;\mathcal F)$ for the 10-point main-text frequency set equals the median, $0.97$, over all $\binom{30}{5}$ sets. We use the same $\mathcal C$ for the fine frequency sweep.
For the interaction-on model, the test accuracy is at least $0.96$ at all 33 points.
For the interaction-off model, the test accuracy is $0.50$ at 8 points.

\begin{table*}[t]
\centering
\small
\begin{tabular}{lcccccc}
\toprule
Model & Number of tasks & Mean & Median & Minimum & \shortstack{Tasks with\\$\Acc=0.50$} & \shortstack{Tasks with\\$\Acc\ge0.95$} \\
\midrule
interaction-on & 33 & 0.9933 & 1.0000 & 0.9600 & 0 & 33 \\
interaction-off & 33 & 0.8736 & 0.9900 & 0.5000 & 8 & 25 \\
\bottomrule
\end{tabular}
\caption{
Summary of the fine frequency sweep for the five regions shown in Fig.~\ref{fig:fine_frequency_selected_subset}.
The accuracy is the empirical accuracy on 100 test samples at each frequency.
The final two columns count the task frequencies satisfying the stated conditions.
}
\label{tab:fine_frequency_selected_subset}
\end{table*}
The results obtained by evaluating all sets selected from the $\Ncand=30$ readout-region samples on the same 33-point task-frequency set, rather than only the $\mathcal C$ used in Fig.~\ref{fig:fine_frequency_selected_subset}, are given in Appendix~\ref{app:qrc_m_dependence}.

\subsection{Proof of the interaction-off proposition}
\label{app:interaction_off_proof}

The test accuracy of $0.50$ at these task frequencies is due not to insufficient classifier capacity but to both classes being mapped to the same feature by the interaction-off feature map. We use the interaction-off model, state preparation, unitary time evolution, pre-readout operation, and time-$T$ observable defined in Sec.~\ref{sec:nv_qrc_protocol}.
We show below, using a $2\times2$ representation of the single-site Hamiltonian, that the interaction-off model has indistinguishable points.
When $H_{\mathrm{dip}}^{(r)}=0$, the time evolution within a region factorizes into a tensor product of site-wise unitaries.
Moreover, the single-site Hamiltonian does not mix $\ket{0}$ with
\begin{equation}
\mathcal H_{\pm}
:=
\operatorname{span}\{\ket{+1},\ket{-1}\}.
\end{equation}
The Hamiltonian for site $i$ in region $r$, restricted to the basis $\{\ket{+1},\ket{-1}\}$, is
\begin{equation}
H_{\pm,i}^{(r)}(t;B)
=
\begin{pmatrix}
D_{\mathrm{gs}}+\gamma_e B(t) & E_{x,i}^{(r)}-\mathrm iE_{y,i}^{(r)}\\
E_{x,i}^{(r)}+\mathrm iE_{y,i}^{(r)} & D_{\mathrm{gs}}-\gamma_e B(t)
\end{pmatrix}.
\label{eq:single_site_pm_hamiltonian}
\end{equation}
Define the single-site time evolution on $\mathcal H_{\pm}$ under the fixed waveform $B_{\phi,\lambda}$ by
\begin{equation}
U_{\pm,i}^{(r)}(T;B_{\phi,\lambda})
:=
\mathcal T
\exp\!\left[
-\mathrm i\int_0^T H_{\pm,i}^{(r)}(t;B_{\phi,\lambda})\,dt
\right]
\end{equation}
In what follows, for any fixed waveform $B(t)$, we use $U_{\pm,i}^{(r)}(T;B)$ to denote the corresponding $2\times2$ unitary on $\mathcal H_{\pm}$.
Absorbing the pre-readout operation into the observable, we define the single-site readout operator as
\begin{equation}
M
:=
U_{\mathrm{prep}}(S_z)^2U_{\mathrm{prep}}^{\dagger}
=
\ket{0}\bra{0}
+
\ket{D}\bra{D}.
\end{equation}
In the following matrix elements, only the restriction of $M$ to $\mathcal H_{\pm}$
\begin{equation}
M_{\pm}
:=
\ket{D}\bra{D}
\end{equation}
contributes.
Because the initial state $\ket{B}$ lies in $\mathcal H_{\pm}$ and the single-site Hamiltonian does not mix $\ket{0}$ with $\mathcal H_{\pm}$, the fixed-waveform feature of the interaction-off model is
\begin{equation}
x_{\Reservoir,\mathrm{off}}^{(r)}(\phi,\lambda)
=
\frac{1}{n}
\sum_{i=1}^{n}
\left|
\bra{D}U_{\pm,i}^{(r)}(T;B_{\phi,\lambda})\ket{B}
\right|^2.
\label{eq:off_feature_site_average}
\end{equation}

For an arbitrary fixed waveform $B(t)$, define
\begin{equation}
B^\sharp(t):=-B(T-t).
\end{equation}
Let $\sigma_x$ denote the Pauli matrix that exchanges $\ket{+1}$ and $\ket{-1}$ in the basis $\{\ket{+1},\ket{-1}\}$. Equation~\eqref{eq:single_site_pm_hamiltonian} then gives
\begin{equation}
H_{\pm,i}^{(r)}(t;B^\sharp)
=
\sigma_x
\left[H_{\pm,i}^{(r)}(T-t;B)\right]^{\mathsf T}
\sigma_x.
\label{eq:off_hamiltonian_transpose_identity}
\end{equation}
By discretizing the time-ordered product and taking the continuum limit, we obtain
\begin{equation}
U_{\pm,i}^{(r)}(T;B^\sharp)
=
\sigma_x
\left[U_{\pm,i}^{(r)}(T;B)\right]^{\mathsf T}
\sigma_x
\label{eq:off_unitary_transpose_identity}
\end{equation}
Here $\sigma_x\ket{B}=\ket{B}$ and $\sigma_x\ket{D}=-\ket{D}$. Moreover, because $\ket{B}$ and $\ket{D}$ are real vectors in the basis $\{\ket{+1},\ket{-1}\}$, any $2\times2$ matrix $U$ satisfies $\bra{D}U^{\mathsf T}\ket{B}=\bra{B}U\ket{D}$.
By unitarity,
\begin{equation}
\left|\bra{D}U_{\pm,i}^{(r)}(T;B^\sharp)\ket{B}\right|^2
=
\left|\bra{D}U_{\pm,i}^{(r)}(T;B)\ket{B}\right|^2.
\end{equation}
Therefore, the site-averaged readout expectation at time $T$ is identical for $B$ and $B^\sharp$ in every region. We now apply this general-waveform result to the binary phase-classification task in the main text.

For the two phase classes considered in this work,
\begin{equation}
\begin{aligned}
B_{0,\lambda}(t)&=s\lambda\sin(\omega t),\\
B_{\pi/2,\lambda}(t)&=s\lambda\cos(\omega t).
\end{aligned}
\end{equation}
When $\omega T=3\pi/2 \pmod{2\pi}$,
\begin{equation}
B_{\pi/2,\lambda}(t)
=
-B_{0,\lambda}(T-t)
\end{equation}
so the equality of the time-$T$ readout expectations for the general waveforms implies that paired events from the two classes with the same $\lambda$ satisfy, in every readout region,
\begin{equation}
x_{\Reservoir,\mathrm{off}}^{(r)}(\phi_1,\lambda)
=
x_{\Reservoir,\mathrm{off}}^{(r)}(\phi_0,\lambda)
\label{eq:off_component_equality_supplement}
\end{equation}
The same equality holds componentwise for the feature vector in Eq.~\eqref{eq:qrc_off_parameter_feature}. For the present $T=10~\mu{\mathrm{s}}$ and the grid in Eq.~\eqref{eq:fine_grid}, this condition occurs at
\begin{equation}
f_{\task}
=
1.375+0.5k~\mathrm{MHz},
\qquad
k=0,\ldots,7.
\end{equation}
For test events from the two classes with the same $\lambda$ and the same $\mathcal C$, the mean absolute difference between the five feature components ranged from $1.3\times10^{-13}$ to $2.1\times10^{-13}$ at these points. This is comparable to the numerical precision of the solver and is consistent with Eq.~\eqref{eq:off_component_equality_supplement}.
This equality is a degeneracy at the feature-map stage.
Because the test data in this work include paired events from the two classes with the same $\lambda$, the test accuracy is $0.50$ at these task frequencies.
For the interaction-on model, the time evolution does not factorize into a tensor product of single-site unitaries because of the intersite coupling, so the preceding proof cannot be applied directly. For the same $\mathcal C$, the numerical test accuracies of the interaction-on model at $f_{\task}=2.375$ and $3.375~\mathrm{MHz}$ are $0.99$ and $1.00$, respectively.

\section{Enumeration over readout-region set sizes}
\label{app:qrc_m_dependence}

In Fig.~\ref{fig:main_results}(c), all sets of $\mR=5$ regions selected from the $\Ncand=30$ numerically generated readout-region samples were evaluated on the 10-point task-frequency set $\mathcal F$ in Eq.~\eqref{eq:tested_grid}.
Here we summarize the enumeration for $\mR=1,\ldots,5$ using the same 30 region samples on the 33-point task-frequency set $\mathcal F_{\mathrm{fine}}$ in Appendix~\ref{app:fine_frequency}. For each $\mR$, we enumerate all sets selected from the $\Ncand=30$ readout-region samples satisfying
\begin{equation}
\mathcal C\subseteq\{1,\ldots,\Ncand\},
\qquad
|\mathcal C|=\mR
\end{equation}
There are $\binom{\Ncand}{\mR}$ such sets. They may share regions and are therefore not independent samples.
The performance of each region set is measured by the minimum test accuracy obtained when the same $\mathcal C$ is used at all task frequencies in $\mathcal F_{\mathrm{fine}}$,
\begin{equation}
\Amin^{q}(\mathcal C;\mathcal F_{\mathrm{fine}})
:=
\min_{f_{\task}\in\mathcal F_{\mathrm{fine}}}
\Acc_{\mathrm{test}}^{q}(f_{\task};\mathcal C).
\label{eq:amin_fine_def}
\end{equation}
For each task frequency, we use the same standardization and ridge classifier as in Sec.~\ref{sec:numerical_setup}, training the classifier on the training features from the event group sharing that unknown frequency and evaluating it on the test features. The region set $\mathcal C$ is kept fixed over all task frequencies in $\mathcal F_{\mathrm{fine}}$; only the classifier coefficients are separately estimated from the training features for each event group.

For each $\mR$, we calculate the fraction, over all $\binom{\Ncand}{\mR}$ sets, of region sets whose $\Amin^{q}(\mathcal C;\mathcal F_{\mathrm{fine}})$ exceeds a specified threshold. Table~\ref{tab:qrc_m_dependence_summary} summarizes the enumeration for the interaction-on model.

\begin{table*}[t]
\centering
\small
\begin{tabular}{ccccccc}
\toprule
$\mR$ & $\binom{30}{\mR}$ & p10 & Median & p90 & \shortstack{Fraction of sets\\with value $\ge0.90$} & \shortstack{Fraction of sets\\with value $\ge0.95$} \\
\midrule
1 & 30     & 0.49 & 0.50 & 0.50 & 0      & 0      \\
2 & 435    & 0.49 & 0.50 & 0.53 & 0      & 0      \\
3 & 4060   & 0.50 & 0.53 & 0.75 & 0.0475 & 0.0089 \\
4 & 27405  & 0.53 & 0.72 & 0.95 & 0.3107 & 0.1477 \\
5 & 142506 & 0.74 & 0.95 & 0.98 & 0.8052 & 0.6349 \\
\bottomrule
\end{tabular}
\caption{Evaluation of the interaction-on model using the same $\mathcal C$ at all task frequencies in the 33-point set $\mathcal F_{\mathrm{fine}}$. The p10 and p90 columns are the 10th and 90th percentiles of $\Amin^{\mathrm{on}}(\mathcal C;\mathcal F_{\mathrm{fine}})$ over all $\binom{30}{\mR}$ sets formed by selecting $\mR$ regions from the 30 numerically generated samples. The last two columns give the fractions of sets for which $\Amin^{\mathrm{on}}(\mathcal C;\mathcal F_{\mathrm{fine}})$ is at least $0.90$ and $0.95$, respectively. Because the number of sets varies with $\mR$ and the sets are not mutually independent, the rows should not be interpreted as estimates based on equal numbers of independent trials.}
\label{tab:qrc_m_dependence_summary}
\end{table*}

For $\mR=1,2$, no set satisfies $\Amin^{\mathrm{on}}(\mathcal C;\mathcal F_{\mathrm{fine}})\ge0.90$; such sets first appear for $\mR=3$. For $\mR=5$, the median of $\Amin^{\mathrm{on}}(\mathcal C;\mathcal F_{\mathrm{fine}})$ is $0.95$, and the fractions of sets with values at least $0.90$ and $0.95$ are $80.52\%$ and $63.49\%$, respectively. For the 30 region samples generated here, these fractions increase with $\mR$. The interaction-off model includes the indistinguishable points established in Proposition~\ref{prop:qrc_off_symmetry_indistinguishable}, so its median and maximum are both $0.50$ for every $\mR$. Since $\binom{30}{\mR}$ varies substantially with $\mR$ and the sets may share regions, these fractions are descriptive statistics for the enumerated sets at each $\mR$.

\begin{figure*}[t]
  \centering
  \includegraphics[width=0.98\linewidth]{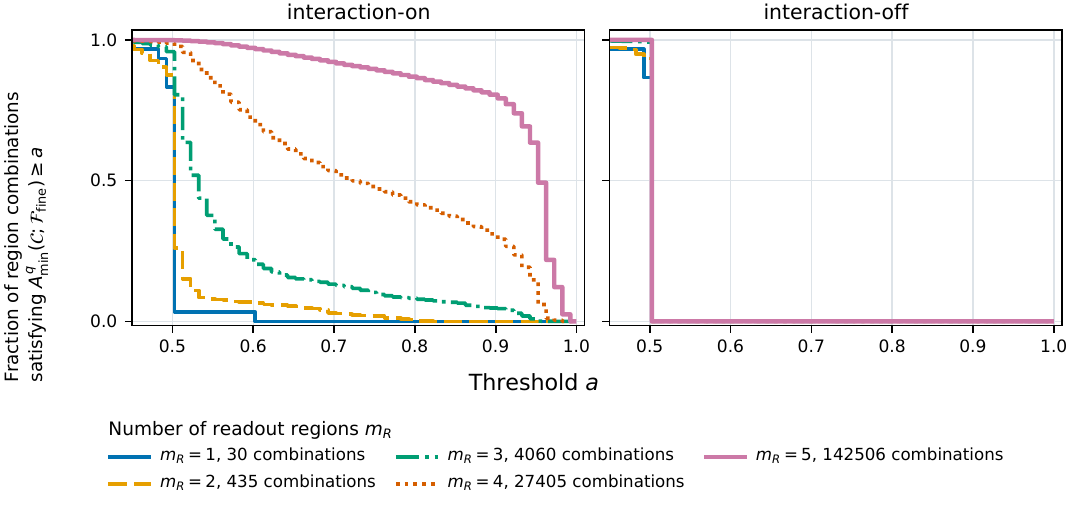}
  \caption{
  Fraction of sets for which $\Amin^{q}(\mathcal C;\mathcal F_{\mathrm{fine}})$ is at least the horizontal-axis value on the 33-point task-frequency set $\mathcal F_{\mathrm{fine}}$.
  The left and right panels show the interaction-on and interaction-off models, respectively.
  $\Amin^{q}(\mathcal C;\mathcal F_{\mathrm{fine}})$ is the minimum test accuracy when the same $\mathcal C$ is used at all 33 points.
  Each curve is obtained by enumerating all $\binom{30}{\mR}$ sets of $\mR$ regions selected from $\Ncand=30$ readout-region samples. The sets may share regions and are therefore not mutually independent.
  }
  \label{fig:qrc_fine_survival_m_dependence}
\end{figure*}

Figure~\ref{fig:qrc_fine_survival_m_dependence} shows, for each threshold $a$ on the horizontal axis, the fraction of region sets satisfying $\Amin^{q}(\mathcal C;\mathcal F_{\mathrm{fine}})\ge a$.
For the interaction-on model, the fraction at each $a$ increases with $\mR$ over the range shown; for $\mR=5$, $80.52\%$ of all region sets satisfy the condition at $a=0.90$.
For the interaction-off model, the indistinguishable points described in Appendix~\ref{app:fine_frequency} imply $\Amin^{\mathrm{off}}(\mathcal C;\mathcal F_{\mathrm{fine}})\le0.50$ for every $\mR$, so no region set satisfies the condition for $a>0.50$.

\FloatBarrier
\section{Applying different CPMG sequences to multiple NV groups}
\label{app:ideal_multi_dd}

In the fixed-DD scheme considered in the main text, one CPMG sequence is applied to all regions, yielding one feature at time $T$.
Here we assume that NV groups with different transition frequencies are assigned to different readout regions and that each group can be controlled independently and frequency-selectively.
Let $m_{\DD}$ be the number of independently applied CPMG sequences, and label the corresponding readout regions by $r=1,\ldots,m_{\DD}$. For the NV group in readout region $r$, set the pulse number to $N_r$ and apply the different CPMG sequences in parallel over the same sensing interval $0\le t\le T$.
Because all pulse numbers are fixed before measurement without using the unknown $f_{\task}$ and are shared by all event groups under evaluation, this setting is also included in fixed-DD\@.
It makes a stronger hardware assumption than the main-text setting because it requires frequency-selective control for each region.

An experiment has demonstrated vector magnetic-field sensing with a Hahn-echo-type pulse measurement by simultaneously controlling NV centers with different crystal axes using microwave pulses containing multiple frequency components~\cite{yahataDemonstrationVectorMagnetic2019}.
The aspect shared with the present section is not the measured quantity or task, but the idea of simultaneously controlling NV groups distinguished by their transition frequencies using a microwave field with multiple frequency components.

For simplicity, we write the equations for $m_{\DD}=2$.
The main-text notation $x_{\DD}(\phi_j,\lambda)$ suppresses the fixed arguments $N_p$ and $\xi$.
To distinguish the multiple pulse numbers in this section, we make $N_1$, $N_2$, and $\xi$ explicit in the arguments of the feature map.
Rather than remeasuring the same magnetic-field waveform with separate pulse sequences, we obtain two features, one from each NV group, during the same sensing interval.

Thus, $\xi:=\omega T$ is common to the two components, while the DD frequency and dimensionless interpulse phase of the CPMG sequence applied to the NV group in readout region $r$ are
\begin{equation}
f_{\DD,r}
:=
\frac{N_r}{2T},
\qquad
\theta_r
:=
\frac{\xi}{N_r}
=
\pi\frac{f_{\task}}{f_{\DD,r}}.
\end{equation}
For a fixed amplitude $\lambda$ and input phase $\phi_j$, define the two-component feature map
\begin{equation}
\bm x_{\DD}(\phi_j,\lambda;\xi,N_1,N_2)
:=
\begin{pmatrix}
\sin[\alpha_j(N_1,\xi)\lambda]\\
\sin[\alpha_j(N_2,\xi)\lambda]
\end{pmatrix}
\label{eq:two_dd_feature}
\end{equation}
The class-conditional feature vector for random amplitude $\Lambda$ is
\begin{equation}
\bm X_{\DD,j}
:=
\bm x_{\DD}(\phi_j,\Lambda;\xi,N_1,N_2)
\end{equation}
Define the coefficient vector by
\begin{equation}
\bm \alpha_{\DD,j}(\xi;N_1,N_2)
:=
\begin{pmatrix}
\alpha_j(N_1,\xi)\\
\alpha_j(N_2,\xi)
\end{pmatrix}.
\end{equation}
If
\begin{equation}
\bm \alpha_{\DD,0}(\xi;N_1,N_2)
=
\bm \alpha_{\DD,1}(\xi;N_1,N_2)
\label{eq:two_dd_coeff_equal}
\end{equation}
holds, then
\begin{equation}
\begin{aligned}
&
\bm x_{\DD}(\phi_0,\lambda;\xi,N_1,N_2)
\\
&\quad=
\bm x_{\DD}(\phi_1,\lambda;\xi,N_1,N_2)
\\
&\hfill
(\lambda\ge0).
\end{aligned}
\end{equation}
Because the random amplitude $\Lambda$ has the same distribution in both classes, their feature distributions coincide. Since the class prior probabilities are equal, any classifier using only this feature vector has accuracy $1/2$, and the Bayes-optimal accuracy is also $1/2$.
This is a sufficient condition for coincident feature distributions that does not rely on a specific property of a particular random-amplitude distribution.
For $m_{\DD}=1$, define the set of $\xi$ values satisfying the coincident-feature condition for pulse number $N$ by
\begin{equation}
\mathcal I_{\DD}(N)
:=
\left\{
\xi>0:
\alpha_0(N,\xi)=\alpha_1(N,\xi)
\right\}
\end{equation}
Then, for $m_{\DD}=2$, the set of $\xi$ values satisfying Eq.~\eqref{eq:two_dd_coeff_equal} is
\begin{equation}
\mathcal I_{\DD}(N_1,N_2)
:=
\mathcal I_{\DD}(N_1)\cap\mathcal I_{\DD}(N_2).
\label{eq:two_dd_indist_set}
\end{equation}
Thus, Eq.~\eqref{eq:two_dd_coeff_equal} does not hold if only one component satisfies $\alpha_0=\alpha_1$ and the other has $\alpha_0\ne\alpha_1$. This fact alone, however, does not imply that the two class-conditional feature distributions differ for a particular random-amplitude distribution.

When $\cos[\xi/(2N_r)]\ne0$, Eqs.~\eqref{eq:alpha0_closed} and~\eqref{eq:alpha1_closed} give, for each component $r=1,2$,
\begin{align}
\alpha_0(N_r,\xi)
&=
g_{N_r}(\xi)
\bigl[
1-\varepsilon_r\cos\xi
\bigr],
\notag\\
\alpha_1(N_r,\xi)
&=
g_{N_r}(\xi)
\varepsilon_r\sin\xi,
\qquad
\varepsilon_r:=(-1)^{N_r}.
\label{eq:two_dd_alpha_components}
\end{align}
Therefore, the condition for $\alpha_0(N_r,\xi)=\alpha_1(N_r,\xi)$ in component $r$ is
\begin{equation}
g_{N_r}(\xi)
\left[
1-\varepsilon_r\cos\xi-\varepsilon_r\sin\xi
\right]
=0.
\label{eq:component_dd_equal_condition}
\end{equation}
For Eq.~\eqref{eq:two_dd_coeff_equal} to hold, both components must satisfy
\begin{equation}
g_{N_r}(\xi)
\left[
1-\varepsilon_r\cos\xi-\varepsilon_r\sin\xi
\right]
=0,
\qquad
r=1,2.
\label{eq:two_dd_equal_condition}
\end{equation}
That is, for each component, either $g_{N_r}(\xi)=0$ makes the responses of both classes zero, or
\begin{equation}
1-\varepsilon_r\cos\xi
=
\varepsilon_r\sin\xi
\label{eq:dd_nonzero_equal_condition}
\end{equation}
must hold. The latter is equivalent to
\begin{equation}
\left(\cos\xi,\sin\xi\right)
=
\left(\varepsilon_r,0\right)
\quad\text{or}\quad
\left(0,\varepsilon_r\right).
\end{equation}
Equation~\eqref{eq:component_dd_equal_condition} applies when Eqs.~\eqref{eq:alpha0_closed} and~\eqref{eq:alpha1_closed} are valid.
At the singular CPMG condition $\xi/N_r=(2p+1)\pi$, the finite sum in Eq.~\eqref{eq:cpmg_finite_sum} must be evaluated directly.
As seen in Eq.~\eqref{eq:dd_resonance_alpha}, this condition generally does not give $\alpha_0=\alpha_1$ for the $\kappa_{\DD}>0$ used here.

If the two pulse numbers have the same parity and both $g_{N_1}(\xi)$ and $g_{N_2}(\xi)$ are nonzero, Eq.~\eqref{eq:two_dd_equal_condition} reduces to the same condition as for a single CPMG sequence.
In contrast, if $\varepsilon_1\ne \varepsilon_2$ and both $g_{N_r}(\xi)$ are nonzero, the two equations
\begin{equation}
1-\varepsilon_1\cos\xi=\varepsilon_1\sin\xi,
\qquad
1-\varepsilon_2\cos\xi=\varepsilon_2\sin\xi
\end{equation}
cannot hold simultaneously.
Thus, combining two CPMG sequences with different parities prevents Eq.~\eqref{eq:two_dd_equal_condition} from holding simultaneously in the regime where both components have nonzero responses.
This conclusion applies only where Eqs.~\eqref{eq:alpha0_closed} and~\eqref{eq:alpha1_closed} are valid; whether equality holds is determined by Eq.~\eqref{eq:two_dd_equal_condition}.
For example, the two-component feature distributions still coincide if one component has $g_{N_r}(\xi)=0$ and the other component satisfies Eq.~\eqref{eq:dd_nonzero_equal_condition} or has $g_{N_r}(\xi)=0$ at the same $\xi$.
In particular, the features of both classes vanish in both components at points satisfying
\begin{equation}
g_{N_1}(\xi)=0,
\qquad
g_{N_2}(\xi)=0.
\label{eq:two_dd_common_zero}
\end{equation}
At these points,
\begin{equation}
\begin{aligned}
\bm x_{\DD}(\phi_0,\lambda;\xi,N_1,N_2)
&=
\bm x_{\DD}(\phi_1,\lambda;\xi,N_1,N_2)\\
&=
\begin{pmatrix}0&0\end{pmatrix}^{\mathsf T},
\qquad \lambda\ge0
\end{aligned}.
\end{equation}
From Eq.~\eqref{eq:cpmg_g_def}, for positive task frequency $f_{\task}$, $g_{N_r}(\xi)=0$ is equivalent to
\begin{equation}
\frac{\xi}{N_r}
=
4\pi \ell_r,
\qquad
\ell_r\in\mathbb Z_{\ge1}.
\label{eq:multi_dd_zero_condition}
\end{equation}
Therefore, the points at which the features of both classes vanish in both components are the discrete points satisfying
\begin{equation}
\xi
=
4\pi \ell_1 N_1
=
4\pi \ell_2 N_2.
\end{equation}

The extension to $m_{\DD}\ge2$ CPMG sequences is analogous.
Let the tuple of pulse numbers be $\bm N:=(N_1,\ldots,N_{m_{\DD}})$.
\begin{equation}
\bm x_{\DD}(\phi_j,\lambda;\xi,\bm N)
:=
\begin{pmatrix}
\sin[\alpha_j(N_1,\xi)\lambda]
\\[-2pt]
\vdots
\\[-2pt]
\sin[\alpha_j(N_{m_{\DD}},\xi)\lambda]
\end{pmatrix}
\end{equation}
At points satisfying
\begin{equation}
\alpha_0(N_r,\xi)
=
\alpha_1(N_r,\xi)
\qquad
(r=1,\ldots,m_{\DD})
\label{eq:mdd_equal_condition}
\end{equation}
the feature vectors satisfy
\begin{equation}
\begin{aligned}
&
\bm x_{\DD}(\phi_0,\lambda;\xi,\bm N)
\\
&\quad=
\bm x_{\DD}(\phi_1,\lambda;\xi,\bm N)
\qquad (\lambda\ge0).
\end{aligned}
\end{equation}
Consequently, the feature distributions of the two classes coincide, and any classifier using only this feature vector has accuracy $1/2$; the Bayes-optimal accuracy is also $1/2$.
Thus, the set of $\xi$ values satisfying Eq.~\eqref{eq:mdd_equal_condition} is
\begin{equation}
\mathcal I_{\DD}(N_1,\ldots,N_{m_{\DD}})
:=
\bigcap_{r=1}^{m_{\DD}}
\mathcal I_{\DD}(N_r).
\end{equation}
As the number of CPMG sequences increases, Eq.~\eqref{eq:mdd_equal_condition} must hold for every component, so the set of $\xi$ values satisfying this sufficient condition may become smaller.
In particular, in the range where Eqs.~\eqref{eq:alpha0_closed} and~\eqref{eq:alpha1_closed} apply and $g_{N_r}(\xi)\ne0$, including pulse numbers of different parity can make the coincident-feature condition harder to satisfy simultaneously across all components.
However, the feature distributions still coincide for $m_{\DD}\ge2$ at points where every component satisfies either Eq.~\eqref{eq:dd_nonzero_equal_condition} or $g_{N_r}(\xi)=0$.
Therefore, when comparing the two class coefficients for multiple features obtained from the NV groups, one should not rely only on Eqs.~\eqref{eq:alpha0_closed} and~\eqref{eq:alpha1_closed}, but return to Eq.~\eqref{eq:cpmg_finite_sum} and compare $\alpha_0(N_r,\xi)$ and $\alpha_1(N_r,\xi)$ directly for each component.

Under this parallel-control setting, different NV groups provide multiple CPMG features during one sensing interval without remeasuring the magnetic-field waveform.
We do not model the fidelity of independent frequency-selective control, the fabrication of frequency-separated readout regions, off-resonant excitation, finite pulse widths, pulse distortions, or region-specific readout.
Thus, the analysis compares the information retained by fixed-DD under this additional control resource with that retained in the main-text setting; it does not propose an experimentally realizable protocol.

\FloatBarrier

\begin{acknowledgments}
D.S. was supported by JST BOOST, Japan, Grant Number JPMJBS2415. We thank Keitaro Anai for useful advice that helped improve the content of this paper. H.K. was supported by the Center of Innovation for Sustainable Quantum AI (JST Grant Number JPMJPF2221).
Y.M. was supported by
JST Moonshot R\&D Grant
Number JPMJMS226C,
JST CREST Grant Number JPMJCR23I5, and Presto
JST Grant Number JPMJPR245B.
\end{acknowledgments}

\bibliographystyle{apsrev4-2}
\bibliography{refs}

\end{document}